\documentclass[journal=tches]{iacrtrans}

\usepackage{url}
\usepackage{hyperref}
\usepackage{xcolor}
\usepackage[nolist]{acronym}
\usepackage{tabularx}
\usepackage{booktabs}
\usepackage{csquotes}
\usepackage{enumitem}
\usepackage{siunitx}
\usepackage{numprint}
\usepackage{xspace}
\usepackage{multirow}
\usepackage{caption}
\usepackage{subcaption}
\usepackage{cleveref}
\usepackage{tikz}
\usepackage{pgfplots}
\usepackage{circledtext}

\definecolor{ibm_blue}{RGB}{100,143,255}
\definecolor{ibm_purple}{RGB}{120,94,240}
\definecolor{ibm_magenta}{RGB}{220,38,127}
\definecolor{ibm_orange}{RGB}{254,97,0}
\definecolor{ibm_yellow}{RGB}{255,176,0}
\definecolor{ibm_green}{RGB}{0,158,115}
\definecolor{ibm_cyan}{RGB}{86,180,233}

\newcommand{\ie}{i.\,e.}
\newcommand{\eg}{e.\,g.}
\newcommand{\cf}{cf.\@\,}

\newcommand{\etal}{et~al.\@\xspace}

\crefname{figure}{Figure}{Figures}
\Crefname{figure}{Figure}{Figures}

\newcolumntype{Y}{>{\centering\arraybackslash}X}
\newcolumntype{L}[1]{>{\raggedright\let\newline\\\arraybackslash\hspace{0pt}}m{#1}}
\newcolumntype{C}[1]{>{\centering\let\newline\\\arraybackslash\hspace{0pt}}m{#1}}
\newcolumntype{R}[1]{>{\raggedleft\let\newline\\\arraybackslash\hspace{0pt}}m{#1}}

\circledtextset{boxfill=black}
\circledtextset{boxtype=o}
\circledtextset{charf=\bfseries\Large}
\circledtextset{charcolor=white}
\circledtextset{charshrink=0.6}
\circledtextset{resize=real}
\circledtextset{width=0.8em}

\newcommand{\BoxOne}{%
  \circledtext[
    boxtype=O,
    boxfill=blue!72.2580645161!cyan!60.7843137255!white,
    boxcolor=blue!72.2580645161!cyan!60.7843137255!white,
    boxlinewidth=0pt
  ]{1}\xspace
}

\newcommand{\BoxTwo}{%
  \circledtext[
    boxtype=O,
    boxfill=red!30.9803921569!yellow,
    boxcolor=red!30.9803921569!yellow,
    boxlinewidth=0pt
  ]{2}\xspace
}

\newcommand{\AppName}{SAMSEM\xspace}

\begin{acronym}
    \acro{API}{application programming interface}

    \acro{BCE}{binary cross entropy}
    
    \acro{DWT}{discrete wavelet transform}

    \acro{ESD}{electrically significant differences}

    \acro{FP}{false positive}
    \acro{FN}{false negative}
    
    \acro{GDSII}{Graphic Data System II}
    \acro{GPU}{graphics processing unit}

    \acro{IC}{integrated circuit}
    \acro{IP}{intellectual property}

    \acro{IoU}{intersection over union}
    
    \acro{MAE}{masked autoencoder}

    \acro{PA}{pixel accuracy}
    
    \acro{SAM}{Segment Anything Model}
    \acro{SAM2}{Segment Anything Model 2}
    \acro{SAM3}{Segment Anything Model 3}
    \acro{SAMIC}{Segment Anything Model for Integrated Circuit Image Analysis}
    \acro{SEM}{scanning electron microscope}

    \acro{TPM}{trusted platform module}
\end{acronym}

\author[C. Gehrmann, J. Ricker, S. Damm, D. Cheng, J. Speith, Y. Shi, A. Fischer, C. Paar]{Christian Gehrmann\inst{1} \and Jonas Ricker\inst{2} \and Simon Damm\inst{2} \and Deruo Cheng\inst{3} \and Julian Speith\inst{1} \and Yiqiong Shi\inst{3} \and Asja Fischer\inst{2} \and Christof Paar\inst{1}}
\institute{
  Max Planck Institute for Security and Privacy (MPI-SP), Bochum, Germany,\\
  \{\email[christian.gehrmann@mpi-sp.org]{christian.gehrmann}, \email[julian.speith@mpi-sp.org]{julian.speith}, \email[christof.paar@mpi-sp.org]{christof.paar}\}@mpi-sp.org
  \and
  Ruhr University Bochum (RUB), Bochum, Germany,\\
  \{\email[jonas.ricker@rub.de]{jonas.ricker}, \email[simon.damm@rub.de]{simon.damm}, \email[asja.fischer@rub.de]{asja.fischer}\}@rub.de
  \and
  Nanyang Technological University (NTU), Singapore, Singapore,\\
  \{\email[deruo.cheng@ntu.edu.sg]{deruo.cheng}, \email[YQSHI@ntu.edu.sg]{yqshi}\}@ntu.edu.sg
}

\title{SAMSEM -- A Generic and Scalable Approach for IC Metal Line Segmentation}

\begin{document}

\maketitle


\begin{abstract}
In light of globalized hardware supply chains, the assurance of hardware components has gained significant interest, particularly in cryptographic applications and high-stakes scenarios.
Identifying metal lines on scanning electron microscope (SEM) images of integrated circuits (ICs) is one essential step in verifying the absence of malicious circuitry in chips manufactured in untrusted environments. 
Due to varying manufacturing processes and technologies, such verification usually requires tuning parameters and algorithms for each target IC.
Often, a machine learning model trained on images of one IC fails to accurately detect metal lines on other ICs.
To address this challenge, we create \AppName by adapting Meta's Segment Anything Model 2 (SAM2) to the domain of IC metal line segmentation.
Specifically, we develop a multi-scale segmentation approach that can handle SEM images of varying sizes, resolutions, and magnifications.
Furthermore, we deploy a topology-based loss alongside pixel-based losses to focus our segmentation on electrical connectivity rather than pixel-level accuracy.
Based on a hyperparameter optimization, we then fine-tune the SAM2 model to obtain a model that generalizes across different technology nodes,  manufacturing materials, sample preparation methods, and SEM imaging technologies.
To this end, we leverage an unprecedented dataset of SEM images obtained from 48 metal layers across 14 different ICs.
When fine-tuned on seven ICs, \AppName achieves an error rate as low as \SI{0.72}{\percent} when evaluated on other images from the same ICs.
For the remaining seven unseen ICs, it still achieves error rates as low as \SI{5.53}{\percent}.
Finally, when fine-tuned on all 14 ICs, we observe an error rate of \SI{0.62}{\percent}.
Hence, \AppName proves to be a reliable tool that significantly advances the frontier in metal line segmentation, a key challenge in post-manufacturing IC verification.
\end{abstract}

\section{Introduction}
\label{mseg::section::introduction}
\Acp{IC} are deployed across all aspects of our digital society.
They provide the foundation not only for our electronic devices, such as smartphones and computers, but also for many safety and security applications in critical infrastructure and defense.
To this end, assurance of the absence of malicious circuitry in such \acp{IC} is essential to ensure the reliability and trustworthiness of their supply chain.
This is particularly true for many cryptographic applications, as hardware Trojans in, \eg, a \ac{TPM}, could easily be leveraged to subvert security altogether.
One means to achieve such assurance is through physical inspection, \ie, destructively opening up \acp{IC}, grinding them down layer by layer while continuously taking images of every layer.
Nowadays, such images are captured using \acp{SEM} due to ever-shrinking \ac{IC} feature sizes in modern \acp{IC}.
These images are then analyzed and compared against a ground truth~\cite{DBLP:conf/sp/PuschnerMBKMP23}, such as \ac{GDSII} design files, or used to extract a gate-level netlist~\cite{DBLP:conf/icdsp/HongCSLG18} for further analysis.

\begin{figure}[htb]
    \centering
    \includegraphics[width=\linewidth]{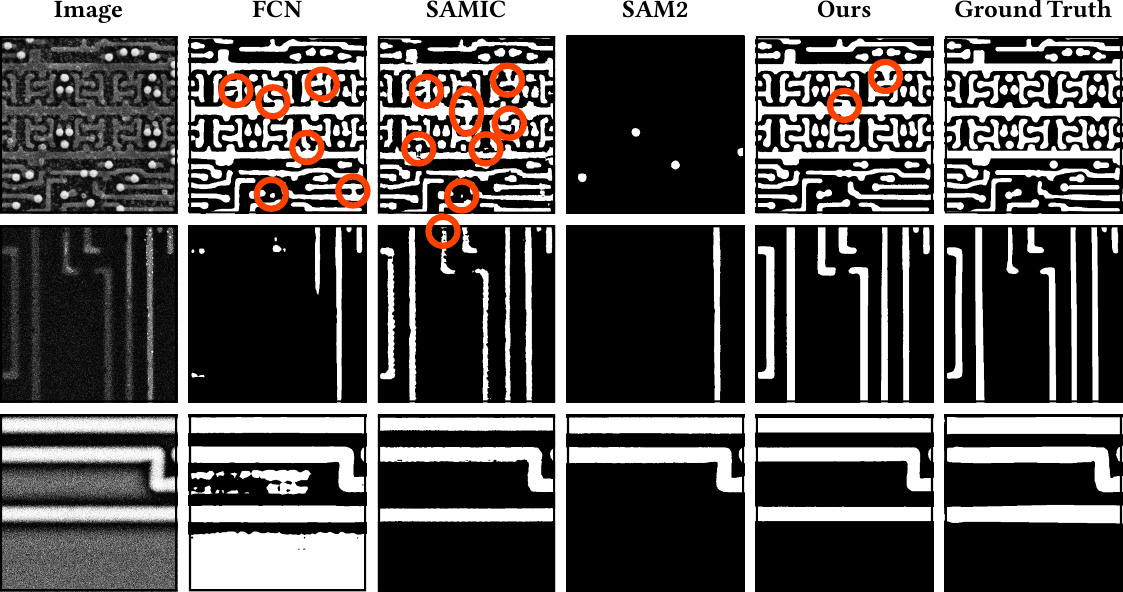}
    \caption{Metal layer images from three different \acp{IC} as well as the corresponding ground-truth mask and the image segmentation masks produced from different methods. Non-obvious \ac{ESD} errors in the segmentation are marked with orange circles.}
    \label{mseg::figure::overview_dataset}
\end{figure}

This destructive imaging process often results in hundreds of gigabytes of images that need to be analyzed in order to ensure the absence of malicious circuitry.
To this end, machine-learning models are increasingly deployed to improve accuracy in image segmentation and recognition tasks while ensuring scalability~\cite{DBLP:conf/ashes/RothaugKABP0P23,DBLP:conf/icdsp/HongCSLG18}.
For the polysilicon layer that implements the transistors of an \ac{IC}, pattern-matching approaches are often deployed~\cite{DBLP:journals/jhss/KimuraSSSESV20,DBLP:conf/isqed/ZhuWDDFW25}. 
In contrast, the metal layers mostly require segmentation of metal lines and vias~\cite{DBLP:journals/jhss/KimuraSSSESV20,DBLP:conf/icdsp/HongCSLG18}. 
However, while this segmentation of typically bright metal lines against a dark background may seem straightforward, it turns out to be rather complicated in practice for multiple reasons:
\begin{enumerate}[label=(\roman*)]
    \item Manufacturing technologies do not only vary between \acp{IC}, but also between different layers of the same \ac{IC}, resulting in vastly different shapes, see \Cref{mseg::figure::overview_dataset} for examples of metal-layer images.
    A segmentation algorithm optimized for one layer of one \ac{IC} may not generalize to other layers or other \acp{IC} without further adjustments.
    \item Delayering errors, artifacts from sample preparation, or dust particles can conceal critical connections and impede even advanced image analysis algorithms.
    \item Structures on modern \ac{IC} are so dense that adjacent but unconnected metal lines may appear to be linked for an automated algorithm due to image distortions or noise, which are inherent to \ac{SEM} images.
    \item The characteristics of \ac{SEM} images differ vastly depending on the employed \ac{SEM}, its detectors, and image capturing settings such as magnification, dwelling time, resolution, and acceleration voltage.
\end{enumerate}

Especially for netlist extraction, a process in which \ac{SEM} images of all \ac{IC} layers are analyzed to recover a gate-level netlist description of the implemented circuit, we can tolerate only very few segmentation errors.
Such errors would significantly impair the analysis of the recovered netlist later on.
Currently, the aforementioned issues are often addressed by manually annotating a subset of the images from each layer of every \ac{IC} and then using these annotated images to train a machine learning model (or adjust parameters in a classical image processing pipeline) to segment the remainder of that layer~\cite{DBLP:conf/ccece/TrindadeUSP18,DBLP:conf/tencon/NgTCG24}, resulting in significant manual overhead.

\paragraph{Our Contributions.}
This work addresses a sub-problem in \ac{IC} image analysis: developing a solution for metal line segmentation that generalizes well across different layers and \acp{IC} without requiring manual intervention or retraining.
This step is essential for reconstructing the interconnections of transistors and, by extension, standard cells on an \ac{IC}, the analysis of which is crucial for hardware assurance.

\Cref{mseg::figure::overview_dataset} depicts segmentation issues observed for techniques for metal line segmentation from the literature when applied to previously unseen \acp{IC}.
To address such generalization issues across metal layers and \acp{IC}, we present \AppName\footnote{\textbf{S}egment \textbf{A}ny \textbf{M}etal-layer \textbf{S}canning \textbf{E}lectron \textbf{M}icroscope image}, which builds on top of Meta's \ac{SAM2}~\cite{DBLP:conf/iclr/RaviGHHR0KRRGMP25} for image segmentation.
While \ac{SAM2} strives for generalized segmentation in natural images, we fine-tune and evaluate the model on metal line segmentation using an unprecedented dataset comprising \ac{SEM} images from 14 different \acp{IC}.
To deal with varying resolutions, magnifications, and metal line structure sizes in \ac{SEM} images, we develop a multi-scale segmentation approach and wrap it around the fine-tuned \ac{SAM2}.

In the context of gate-level netlist extraction, we are less interested in pixel-accurate segmentation than in ensuring correct electrical connectivity.
Therefore, for fine-tuning of the \ac{SAM2} model, we deploy a topology-based loss function in addition to standard pixel-based losses.
In particular, by considering the structure's topology in each image, this loss function explicitly penalizes both short and open connections between metal lines.
This way, we emphasize correct electrical connections over pixel accuracy in the segmentation.
In the context of hardware assurance, this reduces the number of short circuits and open connections observed in an extracted netlist, thereby significantly improving the effectiveness of later analysis steps on the gate-level netlist.
Our experiments are performed on a small cluster of eight Nvidia H100 machine learning accelerators, and fine-tuning on a single Nvidia H100 takes around \SI{69}{\hour} when utilizing the full dataset.
A main advantage of our approach is that, due to SAMSEM's generalization capabilities, fine-tuning is a one-time effort.
Segmentation using SAMSEM can even be executed on consumer-grade hardware.
In summary, our main contributions are:
\begin{itemize}
    \item Throughout our experiments, we utilize an unprecedented dataset of metal-layer images from 14 different \acp{IC}, spanning technology nodes from \SI{200}{nm} down to \SI{20}{nm}.
    \item We develop \AppName by adapting the \ac{SAM2} foundation model to the domain of \ac{IC} metal-layer image segmentation via multi-scale segmentation, a topology-based loss function, and fine-tuning.
    Thereby, we achieve an in-distribution error rate of \SI{0.72}{\percent} when fine-tuning and evaluating on seven different \acp{IC}, compared to \SI{4.44}{\percent} for the best-known approach that we could reproduce.
    \item We demonstrate that \AppName generalizes well, even for unseen \acp{IC}, and therefore does not require retraining when analyzing new \acp{IC}.
    Our approach achieves an out-of-distribution error rate of \SI{5.53}{\percent} on the seven previously unseen \acp{IC}, representing a significant improvement over the \SI{22.25}{\percent} error rate observed for the best-reproduced approach from the literature.
    \item We fine-tune a final model on \SI{90}{\percent} of all images across all 14 \acp{IC}.
    This final model achieves an in-distribution error rate of just \SI{0.62}{\percent}.
\end{itemize}

\paragraph{Data Availability.}
We publish our fine-tuned models, along with all our fine-tuning and benchmarking scripts, as an artifact to this publication, so that others can build upon our work.
For legal reasons, we cannot publish our image datasets.
Images of one of the 14 \acp{IC} used for fine-tuning and evaluation are already publicly available~\cite{DBLP:conf/ashes/RothaugKABP0P23}.
Our artifact can be found at \url{https://doi.org/10.5281/zenodo.21721262}.
\section{Technical Background and Related Work}
\label{mseg::section::background}
In this section, we review relevant technical background and discuss related work on \ac{IC} metal line segmentation in \Cref{mseg::subsection::metal_line_seg}, Meta's \ac{SAM2} in \Cref{mseg::subsection::sam2}, and topology-based loss functions in \Cref{mseg::subsection::topolosses}.

\subsection{Metal Line Segmentation}
\label{mseg::subsection::metal_line_seg}
\Acp{IC} are built from a polysilicon layer at the bottom that (for digital logic) implements transistors and many metal layers on top.
These metal layers establish the electrical connection between the transistors on the polysilicon layer, thereby forming an interconnected circuit.
Hence, a full extraction of all metal lines from \ac{IC} images is an essential step toward hardware assurance.
Machine-learning-based image segmentation algorithms are often employed to this end, which produce a segmentation mask that clearly distinguishes between metal lines and the background.
However, the appearance of metal lines from different \acp{IC}, different layers of the same \ac{IC}, and even under different \ac{SEM} capture settings may vary significantly.
Therefore, metal line segmentation algorithms must often be manually adjusted for the specific dataset on which they operate.

\paragraph{Related Work.}
Trindade \etal propose classical image processing for metal line segmentation~\cite{DBLP:conf/ccece/TrindadeUSP18}, which must be tuned for every new dataset.
Furthermore, they introduce the \ac{ESD} error as a metric to focus on relevant deviations of the segmentation output from the ground truth.
The metric counts the open and short circuits in the segmentation, as well as the \acp{FP} (\ie, segmented metal lines that do not exist) and \acp{FN} (\ie, undetected metal lines in the original images).
This \ac{ESD} metric is more relevant for circuit extraction than pixel-accuracy metrics, as it focuses on errors that result in a faulty extracted circuit description.
Cheng \etal are the first to use machine learning for metal line segmentation by proposing a method based on hybrid K-means clustering and support vector machines~\cite{DBLP:journals/tcas/ChengSLGT18}.
In the same year, Hong \etal introduced the first deep-learning-based technique and evaluated it using a metric similar to the \ac{ESD} metric~\cite{DBLP:conf/icdsp/HongCSLG18}.
Yu \etal built their segmentation solution around HRNet and also used the \ac{ESD} metric for evaluation~\cite{DBLP:conf/icip/YuTGZSTPR22}.
A bit later, Tee \etal concluded that common training approaches do not properly capture the local nature of \ac{ESD} errors~\cite{DBLP:journals/tcasII/TeeHCCSLG23}.
Hence, they put forward a patch-based approach, aiming a discriminator network---comparing the ground truth to intermediate segmentation results---at small patches of the input images.

None of the aforementioned approaches has been tested for whether they generalize to unseen \acp{IC} or even to other layers within the same \acp{IC}.
Instead, they are evaluated on a small dataset often obtained from just a single layer of a single \ac{IC}.
We assume this is the case for economic reasons, as sample preparation and imaging are tedious and costly.
Rothaug \etal approach metal line segmentation using unsupervised learning to reduce the manual workload when facing new datasets~\cite{DBLP:conf/ashes/RothaugKABP0P23,DBLP:journals/jce/RothaugCKABPBP25}.
However, some of their parameters must still be adjusted for every new dataset.
They also used the \ac{ESD} metric for evaluation and released not only their code, but also their training dataset.
To combat manual parameter adjustments altogether, Ng \etal proposed fine-tuning Meta's \ac{SAM}, the predecessor of \ac{SAM2}, using a dataset from four different \acp{IC}, resulting in a model they coined \ac{SAMIC}~\cite{DBLP:conf/tencon/NgTCG24}.
However, they did not evaluate \ac{SAMIC}'s out-of-distribution performance on \acp{IC} not seen during fine-tuning, used a significantly smaller dataset than us, did not employ topological loss functions, and did not address different scales of input images.
In our experiments, we show that \AppName is superior to their approach by around one order of magnitude. 

\subsection{Segment Anything Model 2}
\label{mseg::subsection::sam2}
The \acf{SAM2} is a large foundation segmentation model developed by Meta~\cite{DBLP:conf/iclr/RaviGHHR0KRRGMP25}.
It was primarily designed for a broad range of segmentation tasks on real-world images and videos, but can be fine-tuned for specific applications. 
Compared to other segmentation networks, \ac{SAM2} technically requires a prompt input, which can be a point, a mask, or a bounding box of the object to be segmented.

\begin{figure}[htb]
    \centering
    \includegraphics[width=0.9\linewidth]{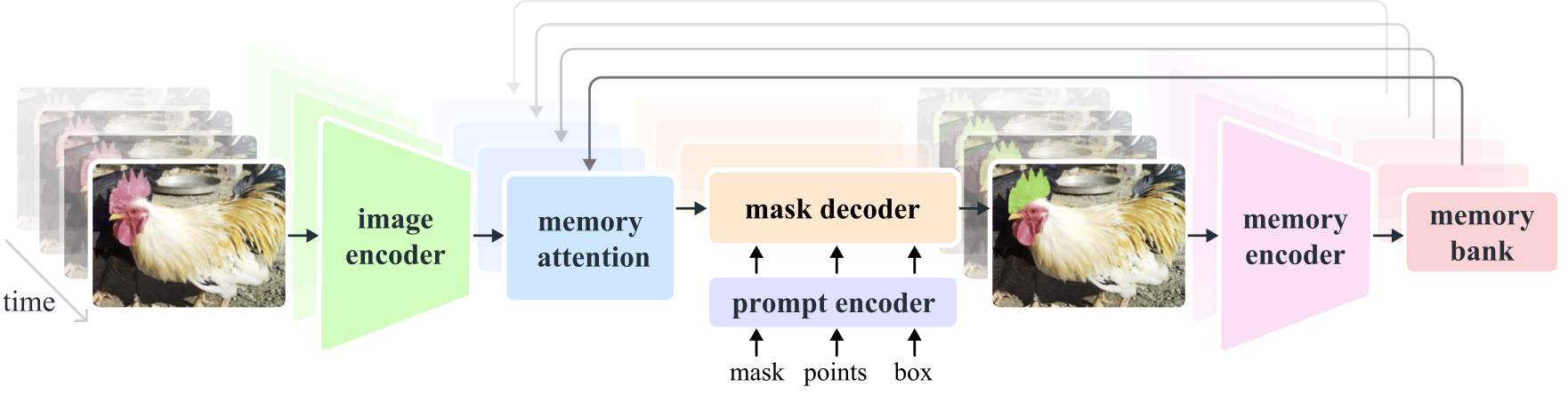}
    \caption{Model components of \ac{SAM2} and their interactions~\cite{DBLP:conf/iclr/RaviGHHR0KRRGMP25}.}
    \label{mseg::figure::sam2}
\end{figure}

\ac{SAM2} comprises an \textit{image encoder}, \textit{memory attention}, \textit{mask decoder}, \textit{prompt encoder}, \textit{memory encoder}, and \textit{memory bank}; see \Cref{mseg::figure::sam2}.
The \textit{image encoder} works in conjunction with the \textit{memory attention} to generate image embeddings for the mask decoder.
The three memory components are used for videos and have no effect on image segmentation; hence, we disregard them. 
For image encoding, a Hiera image encoder~\cite{DBLP:conf/icml/RyaliHB000ACPHM23} is used, pre-trained with a \ac{MAE}~\cite{DBLP:conf/cvpr/HeCXLDG22}.
This image encoder extracts visual features from the input images and provides these feature embeddings as unconditioned tokens to the mask decoder (via the memory attention module). 
Next, the \textit{prompt encoder} accepts a range of user input prompts, such as masks, points, and boxes.
For our work, we focus on input points to generate positional encodings.
The point prompt can be either positive or negative, indicating whether the foreground or background should be selected.
The \textit{mask decoder} generates final segmentation masks from feature embeddings produced by the image encoder and from user prompts processed by the prompt encoder.
While the produced segmentation mask is usually binary, confidence scores for each pixel indicating its likelihood of belonging to the segmented object can also be accessed.

\paragraph{Related Work.}
Numerous works have built upon \ac{SAM2} to improve its performance in specific domains.
For example, Mandal \etal~\cite{DBLP:journals/corr/abs-2510-10288} proposed SAM2LoRA to efficiently fine-tune \ac{SAM2} for retinal fundus segmentation by only considering specific parameters for fine-tuning and freezing the remaining ones.
To better cope with varying image resolutions, Liu \etal~\cite{DBLP:conf/compay/LiuYDPV24} present WSI-SAM, which aims to handle high-resolution whole-slide images in the context of pathology.
Although their approach implies changes to the \ac{SAM} architecture, it does not require full retraining of the entire model.
In a similar vein, Gao \etal~\cite{DBLP:conf/mm/GaoZYL24} improve \ac{SAM} for salient object detection by better capturing multi-scale features and preserving fine-grained details.
HQ-SAM by Ke \etal~\cite{DBLP:conf/nips/KeYDLTT023} adds a learnable high-quality output token to \ac{SAM}, and fuses features from different layers to produce much more accurate and detailed segmentation masks, especially for objects with intricate shapes
Finally, Meta has released the \ac{SAM3}~\cite{carion2025sam} shortly before the submission of our work.
At the time of writing, \ac{SAM3} is available only for fine-tuning on request and is not yet publicly accessible.

\subsection{Topology-Based Loss Functions}
\label{mseg::subsection::topolosses}
Pixel-based loss functions like the Dice loss~\cite{DBLP:conf/miccai/SudreLVOC17} or the \ac{IoU} loss~\cite{DBLP:conf/cvpr/RezatofighiTGS019} are commonly used for training image-related machine learning models.
These loss functions penalize pixel differences between the model output (\eg, a segmentation mask) and the ground truth.
In contrast to pixel-based loss functions, topology-based loss functions prioritize topological accuracy over pixel-level accuracy.
Such loss functions are often researched in the domain of biological cell boundary detection and automated street recognition in satellite images~\cite{DBLP:conf/nips/HuLSC19,DBLP:conf/iclr/HuWLSC21}.
In our case, we are interested in correct electrical connections rather than the accurate segmentation of every pixel on an \ac{IC} image.
Hence, topology-based loss functions seem to be a natural fit.

\paragraph{Related Work.}
In their seminal work on topology-based image segmentation, Hu \etal propose a loss function based on Betti numbers that count structures such as the connected components and holes in an input image~\cite{DBLP:conf/nips/HuLSC19}.
They use persistence diagrams to measure the topological similarity between the ground truth and the model prediction.
Later, Shit \etal proposed clDice as a loss function specifically designed for tubular structures such as vascular networks and roads~\cite{DBLP:conf/cvpr/ShitPSEUZPBM21}. 
The loss compares the centerline of a predicted tubular structure (\ie, its skeleton) to the ground truth.
Liu \etal later enhanced a skeleton-based loss function with better awareness of structure boundaries~\cite{DBLP:conf/ijcai/LiuMBXWXM024}.

The DMT loss proposed by Hu \etal uses discrete Morse theory to identify unwanted critical points in the predicted segmentation~\cite{DBLP:conf/iclr/HuWLSC21}.
Similar to Hu \etal~\cite{DBLP:conf/nips/HuLSC19}, persistence diagrams of the prediction are compared to the ground truth.
Hu \etal further advance the field by using homotopy warping to consider not only the number of topological features, but also their geometric placement~\cite{DBLP:conf/nips/Hu22}.
It identifies topologically critical pixels by warping the predicted mask into the ground truth.

Stucki \etal built upon the initial publication of Hu \etal~\cite{DBLP:conf/nips/HuLSC19}.
Instead of simply measuring overall topological counts (\ie, Betti numbers), it matches individual topological features between the predicted segmentation and the ground truth in a spatially consistent way, defining a Betti matching error that can be used as a differentiable loss to improve the segmentation's topological correctness~\cite{DBLP:conf/icml/StuckiPSMB23}.
Similarly, Wen \etal encode spatial awareness with persistence diagrams to construct their loss function~\cite{DBLP:journals/corr/abs-2412-02076}.
Expanding on their own work, Stucki \etal adjust Betti matching to 3D segmentation by developing a faster implementation of the Betti matching loss, making topology-based loss functions practical for volumetric data~\cite{DBLP:journals/corr/abs-2407-04683}.
Berger \etal then extend Betti matching to multi-class segmentation by reducing an $n$-class segmentation to $n$ single-class segmentation tasks and applying induced barcode matchings for each class~\cite{DBLP:conf/miccai/BergerLSBSBRBP24}.

Finally, Topograph by Lux \etal encodes the topology as a component graph, with nodes corresponding to pixels or regions thereof and edges representing adjacency~\cite{DBLP:conf/iclr/LuxBWSRBP25}.
By building their loss functions solely on graph algorithms, they achieve better efficiency than most other algorithms while providing strong topological guarantees for segmentation.

\section{Methodology}
\label{mseg::section::method}
We focus on segmenting metal lines while ignoring vias, since vias can be reliably detected using thresholding or machine learning~\cite{DBLP:conf/icpr/SinglaLG20}.
In this section, we describe how we construct \AppName to produce reliable segmentation results, even for unseen \acp{IC}.
After motivating fine-tuning in \Cref{mseg::subsection::motivation} and introducing our multi-scale segmentation pipeline in \Cref{mseg::subsection::seg_pipeline}, we describe our dataset comprising 14 \acp{IC} in \Cref{mseg::subsection::dataset}.
Next, in \Cref{mseg::subsection::training}, we explain our fine-tuning, data augmentation, and (topological) loss function.
We then report the best-suited parameters for our segmentation pipeline as determined through hyperparameter optimization in \Cref{mseg::subsubsection::hyperparam}.
Finally, we describe how we generate segmentation masks with \AppName in \Cref{mseg::subsection::inference}.

\subsection{Motivation}
\label{mseg::subsection::motivation}
\ac{SAM2} was trained on a large and diverse dataset of natural images.
It was designed to segment RGB images of objects, animals, or people. 
\AppName is based on \ac{SAM2}, but targets the non-interactive segmentation of metal layer images from \acp{IC} taken with \acp{SEM}.
Our \ac{IC} images differ significantly from the natural images in that they are grayscale and exhibit repeating structures, but little variation in their features.
When simply segmenting on our metal layer images, \ac{SAM2} (without fine-tuning) produces catastrophic results with a \ac{PA} of only 0.639 and an \ac{ESD} error rate of \SI{75.6}{\percent}, see \Cref{mseg::table::model_components}.
To truly leverage the power of \ac{SAM2} for \ac{IC} metal layer image segmentation, we must adapt the model to our specific domain and fine-tune the \ac{SAM2} model on representative data.

Given SAMSEM's expected generalization performance on unseen \acp{IC}, we consider fine-tuning a one-time effort. 
Our goal is that, after SAMSEM has been fine-tuned on \ac{IC} metal-layer images once, it can be used to segment unseen \acp{IC} without further adaptation.
In contrast to fine-tuning, which we perform on Nvidia H100 accelerators, segmentation could even be performed on consumer hardware.
Given the complex sample preparation and imaging steps preceding \ac{IC} image segmentation, which require expensive tools and can take weeks to months to complete, we argue that even the overhead of fine-tuning would have no significant impact on the overall effort required for full netlist extraction.

\subsection{Segmentation Pipeline}
\label{mseg::subsection::seg_pipeline}
We now discuss the segmentation pipeline of our metal-layer segmentation system based on \ac{SAM2}.
The pipeline accounts for the fact that \ac{SEM} images of metal layers may differ significantly in shape, size, and resolution.

\subsubsection{Model \BoxOne\ -- Segmenting Original-Size Images}
\label{mseg::subsubsection::model_one}
\begin{figure}[htbp]
  \centering
  \begin{subfigure}{0.3\linewidth}
    \centering
    \includegraphics[width=\linewidth]{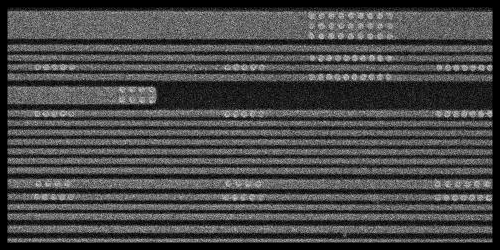}
    \caption{Input image.}
    \label{mseg::subfigure::large_error_input}
  \end{subfigure}
  \hfill
  \begin{subfigure}{0.3\linewidth}
    \centering
    \includegraphics[width=\linewidth]{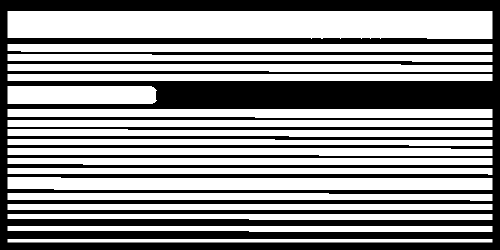}
    \caption{Ground truth.}
    \label{mseg::subfigure::large_error_ground_truth}
  \end{subfigure}
  \hfill
  \begin{subfigure}{0.3\linewidth}
    \centering
    \includegraphics[width=\linewidth]{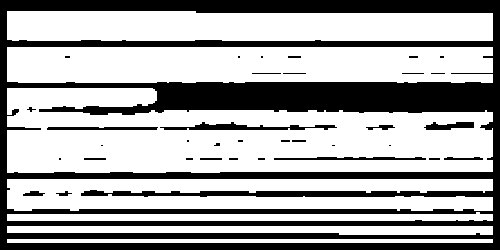}
    \caption{Segmentation mask.}
    \label{mseg::subfigure::large_error_segmentation}
  \end{subfigure}

  \caption{\circledtext[boxfill=red,boxcolor=red]{A} -- (\subref{mseg::subfigure::large_error_input}) to (\subref{mseg::subfigure::large_error_segmentation}) depict short circuits in the segmentation mask resulting from downscaling the input image to fit the shape expected by the image encoder.}
  \label{mseg::figure::large_error}
\end{figure}

Initially, we fine-tuned \ac{SAM2} by simply providing the original-size images obtained from an \ac{SEM}, along with the ground truth.
We refer to this model as model~\BoxOne.
The \ac{SAM2} image encoder operates on $1024 \times 1024$-pixel images; hence, images of other sizes are simply scaled to fit its resolution.
In our setting, we often encounter high-resolution images of $10,000 \times 10,000$ pixels or more, as these are typical output sizes for \acp{SEM}.
When fed to the image encoder, these images are simply scaled down to $1024 \times 1024$ pixels.
However, such metal layer images often depict fine-grained structures that are then brought closer together (in terms of pixel distance) or even fused during downscaling.
This, in turn, \circledtext[boxfill=red,boxcolor=red]{A} causes short circuits in the segmentation when compared to the ground truth, thereby significantly increasing the \ac{ESD} error rate; see the fused metal lines in \Cref{mseg::subfigure::large_error_segmentation}.

\subsubsection{Model \BoxTwo\ -- Segmenting Images in Smaller Patches}
\label{mseg::subsubsection::model_two}
A straightforward approach to address this issue would be to simply cut the input images into smaller patches and feed only these patches into the \ac{SAM2} image encoder for segmentation.
To this end, we fine-tuned a second model, named model~\BoxTwo, on $512 \times 512$-pixel patches only.
The patches are cut out with at least \SI{10}{\percent} overlap and upscaled to $1024 \times 1024$ pixels to fit the image encoder shape.
The upscaling proved useful as it increases the pixel distance between closely neighboring metal lines. 

\begin{figure}[htbp]
  \centering
  \begin{subfigure}{0.3\linewidth}
    \centering
    \includegraphics[width=\linewidth]{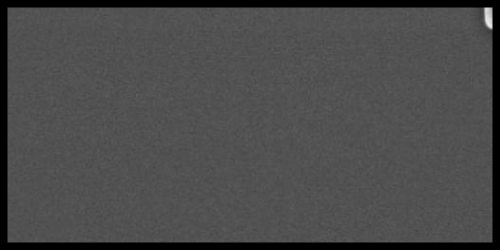}
    \caption{Input image.}
    \label{mseg::subfigure::inverse_error_input}
  \end{subfigure}
  \hfill 
  \begin{subfigure}{0.3\linewidth}
    \centering
    \includegraphics[width=\linewidth]{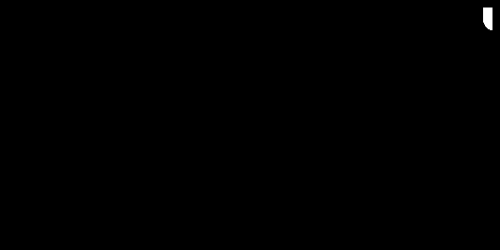}
    \caption{Ground truth.}
    \label{mseg::subfigure::inverse_error_ground_truth}
  \end{subfigure}
  \hfill
  \begin{subfigure}{0.3\linewidth}
    \centering
    \includegraphics[width=\linewidth]{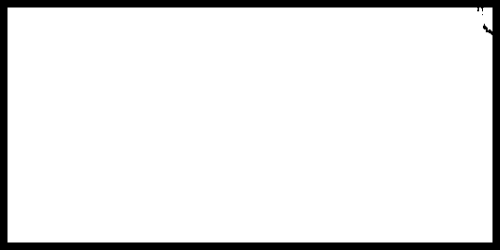}
    \caption{Segmentation mask.}
    \label{mseg::subfigure::inverse_error_segmentation}
  \end{subfigure}\\\vspace{.5cm}

  \begin{subfigure}{0.3\linewidth}
    \centering
    \includegraphics[width=\linewidth]{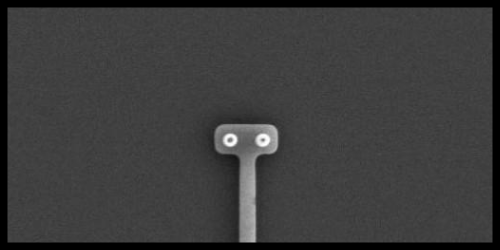}
    \caption{Input image.}
    \label{mseg::subfigure::flaky_error_input}
  \end{subfigure}
  \hfill
  \begin{subfigure}{0.3\linewidth}
    \centering
    \includegraphics[width=\linewidth]{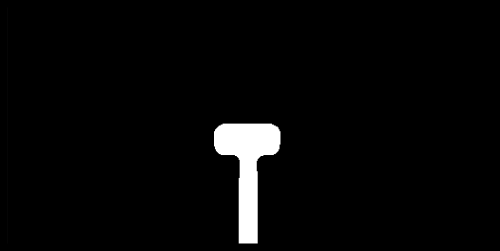}
    \caption{Ground truth.}
    \label{mseg::subfigure::flaky_error_ground_truth}
  \end{subfigure}
  \hfill
  \begin{subfigure}{0.3\linewidth}
    \centering
    \includegraphics[width=\linewidth]{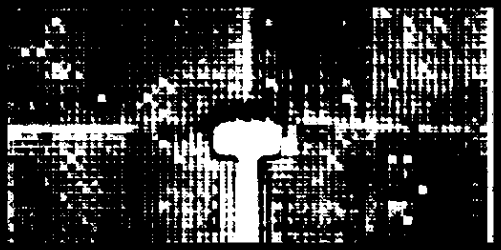}
    \caption{Segmentation mask.}
    \label{mseg::subfigure::flaky_error_segmentation}
  \end{subfigure}

  \caption{\circledtext[boxfill=red,boxcolor=red]{B} -- (\subref{mseg::subfigure::inverse_error_input}) to (\subref{mseg::subfigure::inverse_error_segmentation}) depict background being segmented as a metal line because of the lack of structures in the input image, resulting in \acp{FP} or short circuit.
  \circledtext[boxfill=red,boxcolor=red]{C} -- (\subref{mseg::subfigure::flaky_error_input}) to (\subref{mseg::subfigure::flaky_error_segmentation}) show white speckles around the correctly identified metal line in the segmentation mask, resulting in vast amounts of false positives.}
  \label{mseg::figure::small_error}
\end{figure}

Model \BoxTwo significantly improves upon the issue observed in model \BoxOne and thereby reduces the number of shorts in the segmentation.
However, it also generates a different class of errors that drastically increases the number of false positives in the resulting segmentation mask.
Analyzing these errors, we found some input image patches that comprise primarily, or even entirely, of either background or foreground structures.
For these patches, our model~\BoxTwo hallucinates segmentations that do not exist in the actual input images.
Here, we observed two types of errors stemming from these hallucinations: Firstly, \circledtext[boxfill=red,boxcolor=red]{B} the segmentation mask may end up being mostly inverse to the ground truth, \eg, background is segmented as a metal line, see 
\cref{mseg::subfigure::inverse_error_input,mseg::subfigure::inverse_error_ground_truth,mseg::subfigure::inverse_error_segmentation}.
Secondly, \circledtext[boxfill=red,boxcolor=red]{C} parts of the patch may be wrongly classified as a metal line with speckled transitions to background, see \cref{mseg::subfigure::flaky_error_input,mseg::subfigure::flaky_error_ground_truth,mseg::subfigure::flaky_error_segmentation}.
Here, each white speckle would be interpreted as a false positive by the \ac{ESD} metric, thereby vastly increasing our error rate.

\subsubsection{Multi-Scale Segmentation Approach}
\label{mseg::subsubsection::multi_scale}
\begin{figure}[htbp]
    \centering
    \includegraphics[width=\linewidth]{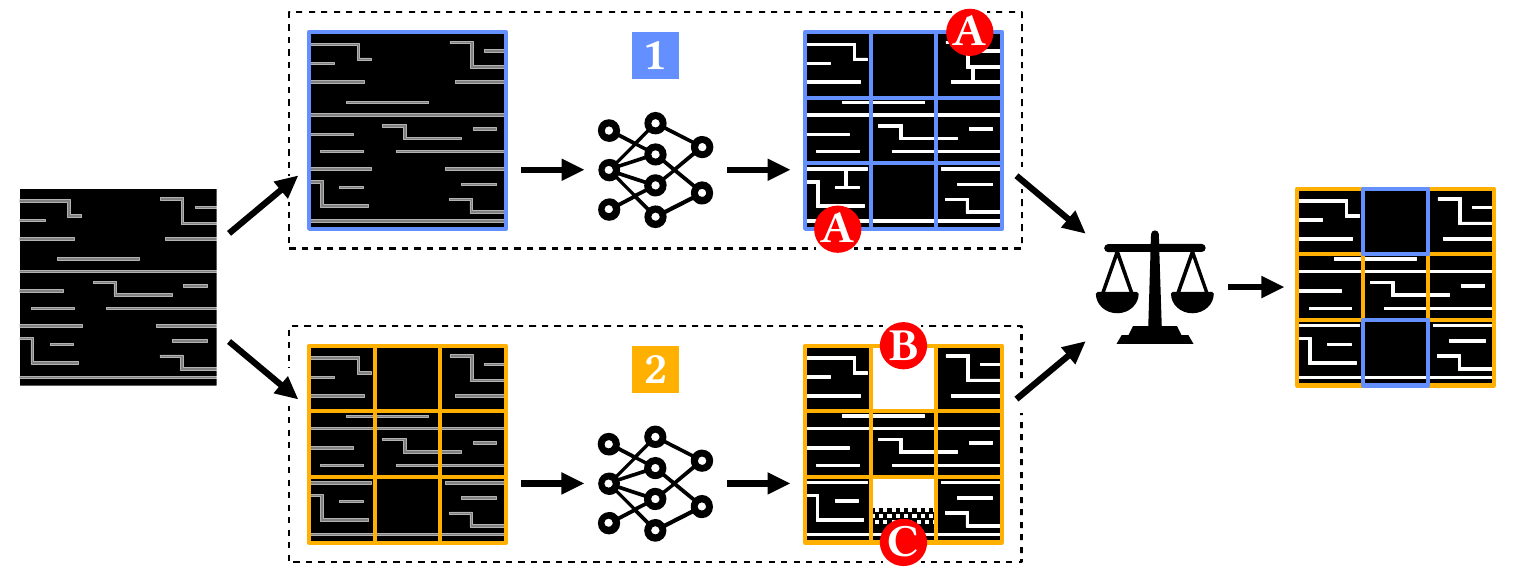}
    \caption{Workflow of our multi-scale segmentation approach.}
    \label{mseg::figure::two_model}
\end{figure}

To address both issues at once, we introduce a multi-scale segmentation approach, as depicted in \Cref{mseg::figure::two_model}. 
We simultaneously segment the original-size image using model~\BoxOne and $512 \times 512$-pixel patches extracted from the input image using model~\BoxTwo.
This multi-scale approach ensures that our fully-automated segmentation algorithm performs well for both small and large structures in arbitrary metal layer images and operates reliably, independent of a \ac{SEM}'s magnification and resolution.
By working on full-size images, we ensure that \ac{SAM2} is provided sufficient context for larger structures that extend beyond a single patch, while working on patches improves segmentation performance for fine-grained structures.
This approach produces two segmentation masks for each input image: one composed of $512 \times 512$-pixel patches and one directly corresponding to the segmentation of the original-sized image.

Both masks are analyzed patch-by-patch to construct the final segmentation mask.
Since \ac{SAM2} always produces segmentation masks of the size of the input image, the mask produced by model~\BoxOne is cut into patches of $512 \times 512$ pixels.
The masks produced by model~\BoxTwo are already of size $512 \times 512$.
Our decision algorithm, which we present hereafter, then composes the final segmentation mask by choosing patches from either model~\BoxOne or model~\BoxTwo based on lightweight quality metrics.
This choice is made for each segmentation mask patch, so the final mask can comprise patches from both approaches.
Simply put, the decision algorithm favors segmentation mask patches from model~\BoxTwo for fine-grained structures in images, while selecting patches from model~\BoxOne for images depicting larger structures or lacking structure altogether.
Thereby, we combine the advantages of both approaches while mitigating the issues depicted in Figures~\ref{mseg::figure::large_error} and \ref{mseg::figure::small_error}.

\subsubsection{Deciding Between Patches from Model \BoxOne or Model \BoxTwo}
Our decision algorithm, which chooses between patches from models~\BoxOne and~\BoxTwo, is based on classical computer vision techniques applied to the segmentation mask patches produced by either approach.
Given that the errors \circledtext[boxfill=red,boxcolor=red]{B} and \circledtext[boxfill=red,boxcolor=red]{C} always produce reproducible patterns, detecting their presence is straightforward.
To this end, we count the small components within a segmentation mask patch to determine the number of speckles in the segmentation.
Our experiments have shown that many such speckles indicate a noisy segmentation that should be disregarded.
For a detected component to be counted as a speckle, it must be at most 16 pixels in size, because larger components are usually valid segmentations.

\begin{figure}[htbp]
    \centering
    \includegraphics[width=0.7\linewidth]{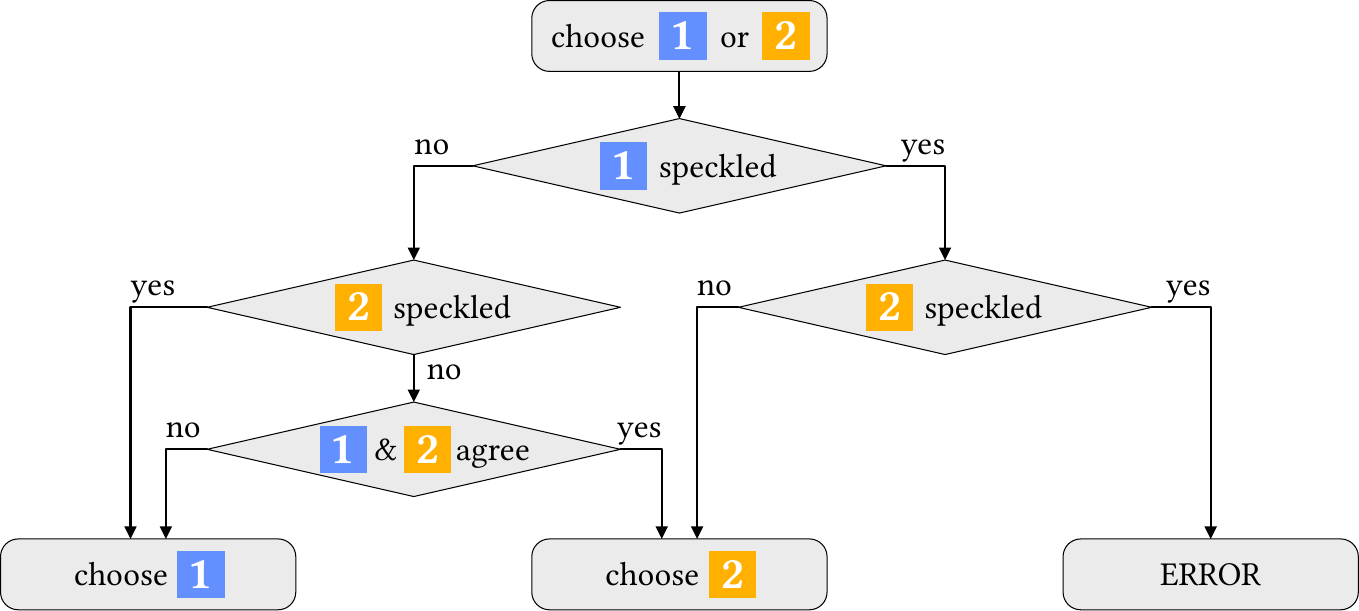}
    \caption{Decision diagram illustrating how patches from either model \BoxOne or \BoxTwo are chosen, or an error is produced, based on their segmentation quality.}
    \label{mseg::figure::patch_decision}
\end{figure}

\Cref{mseg::figure::patch_decision} depicts the decision diagram for how a segmentation mask patch from models~\BoxOne or model~\BoxTwo is selected.
Here, the property \enquote{speckled} refers to a patch containing at least 50 speckles for model~\BoxOne and 50 speckles for model~\BoxTwo.
Furthermore, two patches (one from each model) \enquote{agree} if at least \SI{60}{\percent} of their pixels are identical.
These thresholds were determined through a grid search in which we automatically tested different parameter values to minimize the number of \ac{ESD} errors across segmentation patches.
If both the patch from model~\BoxOne and the one from model~\BoxTwo are considered \enquote{speckled}, an error is reported, and the respective patch of the original image is flagged for manual inspection.

\subsection{Datasets}
\label{mseg::subsection::dataset}
Our vast and diverse dataset contains images from 48 metal layers of 14 different \acp{IC}.
To the best of our knowledge, previous work has operated on images of at most four \acp{IC}, \cf Ng \etal~\cite{DBLP:conf/tencon/NgTCG24}.
Our dataset includes these four \acp{IC}, one \ac{IC} published by Rothaug \etal~\cite{DBLP:conf/ashes/RothaugKABP0P23}, and multiple layers from 9 additional \acp{IC} captured by our team.
While details on these \acp{IC} cannot be disclosed for legal reasons, they range from around \SI{200}{nm} down to \SI{20}{nm} structure size and are composed of varying materials.
The metal layer images were captured using different \acp{SEM}, different capturing settings, and also different sample preparation techniques.
As shown in \Cref{mseg::figure::overview_dataset}, the resulting dataset is quite diverse in the shapes, numbers, and sizes of the metal lines.
This dataset diversity is essential for fine-tuning the \ac{SAM2} model so that it performs well even on \acp{IC} it has not seen during fine-tuning. 
Finally, the dataset enables us to effectively test both the in-distribution performance of \AppName on \acp{IC} it has seen during fine-tuning and its out-of-distribution performance on unseen \acp{IC}, see \Cref{mseg::section::eval}.

\subsubsection{Dataset Composition}
Overall, a small fraction of the metal layer images captured from each \ac{IC} has been semi-automatically annotated and manually verified to generate a ground truth for training, validation, and testing.
Hence, each image in our dataset is accompanied by a corresponding binary ground-truth mask.
\Cref{mseg::table::dataset} lists the absolute number of annotated images for each \ac{IC}, the number of $512 \times 512$ pixel image patches, and the number of metal layers from which these images have been taken.
Due to the variety of capture techniques used for the different \acp{IC}, the full-size images vary widely in size.
Hence, the number of patches is better suited for a comparison in terms of the amount of available data per \ac{IC}.
For \acp{IC} 8 to 14, rather few images have been annotated.

\begin{table}[htb]
    \centering
    \footnotesize
    \begin{tabularx}{\linewidth}{XC{1.2cm}C{1.2cm}C{1.2cm}C{1.2cm}C{1.2cm}C{1.2cm}C{1.2cm}C{1.2cm}}
        \toprule
        \textbf{IC} & 1 & 2 & 3 & 4 & 5 & 6 & 7\\
        \midrule
        \textbf{\# images} & 150 & 70 & 90 & 50 & 50 & 140 & 321\\
        \textbf{\# patches} & \numprint{4050} & \numprint{1120} & \numprint{1950} & \numprint{1000} & \numprint{1000} & \numprint{2600} & \numprint{23112}\\
        \textbf{\# layers} & 3 & 7 & 2 & 1 & 4 & 3 & 1\\
        \midrule\midrule
        \textbf{IC} & 8 & 9 & 10 & 11 & 12 & 13 & 14\\
        \midrule
        \textbf{\# images} & 12 & 5 & 5 & 3 & 3 & 28 & 10\\
        \textbf{\# patches} & 756 & 342 & 254 & 189 & 48 & \numprint{2735} & 810\\
        \textbf{\# layers} & 4 & 5 & 5 & 3 & 3 & 6 & 1\\
        \bottomrule
    \end{tabularx}
    \caption{Number of images available from each \ac{IC}, number of $512 \times 512$ pixel patches, and number of metal layers from which the images were taken.}
    \label{mseg::table::dataset}
\end{table}

\subsubsection{Dataset Splits}
For fine-tuning (see \Cref{mseg::subsection::training}), the hyperparameter optimization (see \Cref{mseg::subsubsection::hyperparam}), and evaluation (see \Cref{mseg::section::eval}), we split the original-size images from \acp{IC} 1 to 7 into three distinct datasets: \SI{70}{\percent} of the images of each \ac{IC} are used for training, \SI{10}{\percent} for validation and model selection during the hyperparameter search, and \SI{20}{\percent} for testing and evaluation of the in-distribution segmentation accuracy on \acp{IC} that the model has seen during fine-tuning.
To evaluate the out-of-distribution performance of \AppName on unseen \acp{IC}, all images of \acp{IC} 8 to 14 are used for testing.
In particular, none of the images from \acp{IC} 8 to 14 are used for training, validation, or model selection.
In a separate experiment, we use \SI{90}{\percent} of all images from all 14 \acp{IC} to fine-tune our final model, using the hyperparameters determined beforehand (see \Cref{mseg::subsection::full_data}). 
In this case, the remaining \SI{10}{\percent} of the images are used to evaluate the segmentation accuracy of this final model.

\subsection{Fine-Tuning}
\label{mseg::subsection::training}
In this section, we provide details on how we fine-tune \ac{SAM2} for \ac{IC} metal layer image segmentation.
In particular, we discuss the need for fine-tuning, the data augmentation techniques we deployed, and our choice of loss functions.

\subsubsection{Selection of Model Components}
For \ac{SAMIC}~\cite{DBLP:conf/tencon/NgTCG24}, Ng \etal only fine-tuned the mask decoder of \ac{SAM} (version~1).
However, we argue that to thoroughly investigate the potential of \ac{SAM2} for metal line segmentation, we also need to consider the two remaining model components relevant to image segmentation, \cf \Cref{mseg::subsection::sam2}.
Hence, we now investigate the individual impact of the mask decoder, prompt encoder, and image encoder on \AppName's segmentation performance.
To this end, we fine-tune seven \ac{SAM2} models with all possible combinations of these three components, and compare them against the off-the-shelf \ac{SAM2} model as trained by Meta in \Cref{mseg::table::model_components}.
The evaluation setup is equivalent to the in-distribution tests performed later in \Cref{mseg::subsection::segperf}, \ie, we fine-tune on the training set comprising \SI{70}{\percent} of the images from \acp{IC} 1 to 7 and test on the test set containing \SI{20}{\percent} of the images from the same \acp{IC}.
We compare the impact of the components using standard pixel-based \acf{PA}, Dice~\cite{DBLP:conf/icpr/SinglaLG20}, and \acs{IoU}~\cite{DBLP:conf/cvpr/RezatofighiTGS019} metrics, as well as the \ac{ESD} error rate.
The \ac{ESD} error rate is given as the relative number of \ac{ESD} errors per metal line.

\begin{table}[htb]
    \centering
    \footnotesize
    \begin{tabular}{C{1.2cm}C{1.2cm}C{1.2cm} C{1.2cm}C{1.2cm}C{1.2cm}C{1.2cm}C{1.6cm}}
      \toprule
      \multicolumn{3}{c}{\textbf{fine-tuned components}}
        & \multirow{3}{*}{\textbf{PA} $\uparrow$}
        & \multirow{3}{*}{\textbf{Dice} $\uparrow$}
        & \multirow{3}{*}{\textbf{IoU} $\uparrow$}
        & \multirow{3}{*}{\shortstack{\textbf{ESD} $\downarrow$\\(\%)}}
        & \multirow{3}{*}{\shortstack{\textbf{training}\\\textbf{time}}}\\
      mask & prompt & image &&&&\\
      decoder & encoder & encoder &&&&\\
      \midrule
      \(\square\) & \(\square\) & \(\square\)  & 0.639 & 0.111 & 0.067 & 75.6 & -\\
      \(\blacksquare\) & \(\square\) & \(\square\) & 0.948 & 0.933 & 0.885 & 12.1 & \SI{43}{h}\\
      \(\square\) & \(\blacksquare\) & \(\square\) & 0.673 & 0.332 & 0.236 & 66.1 & \SI{39}{h}\\
      \(\blacksquare\) & \(\blacksquare\) & \(\square\) & 0.944 & 0.927 & 0.879 & 11.5 & \SI{34}{h}\\
      \(\square\) & \(\square\) & \(\blacksquare\) & \underline{0.972} & \underline{0.960} & 0.924 & 0.8 & \SI{51}{h}\\
      \(\blacksquare\) & \(\square\) & \(\blacksquare\) & 0.971 & 0.957 & 0.921 & 1.0 & \SI{42}{h}\\
      \(\square\) & \(\blacksquare\) & \(\blacksquare\) & 0.969 & 0.954 & 0.917 & 1.4 & \SI{43}{h}\\
      \(\blacksquare\) & \(\blacksquare\) & \(\blacksquare\) & \underline{0.972} & \underline{0.960} & \underline{0.925} & \underline{0.7} & \SI{44}{h}\\
      \bottomrule
    \end{tabular}

    \caption{Ablation study for \AppName's performance when fine-tuning only selected components of \ac{SAM2}. 
    The first line refers to the default \ac{SAM2} model without fine-tuning.}
    \label{mseg::table::model_components}
\end{table}

From \Cref{mseg::table::model_components}, we conclude that the image encoder has the most impact on the segmentation accuracy, achieving an \ac{ESD} error rate of \SI{0.8}{\percent}.
Fine-tuning the mask decoder also significantly reduces the error rate to \SI{12.1}{\percent}.
The prompt encoder has little impact on segmentation and still results in a \SI{66.1}{\percent} \ac{ESD} error rate.
Combining different encoders and decoders naturally produces better results than evaluating them in isolation.
We also see that combining all three components yields the lowest \ac{ESD} error rate of \SI{0.7}{\percent} and the best pixel accuracy, although this is only a marginal improvement over fine-tuning the image encoder alone.
Still, given that there is no significant overhead in training time when combining all three approaches compared to fine-tuning the image encoder alone, we decided to fine-tune the mask decoder, prompt encoder, and image encoder together.

\subsubsection{Fine-Tuning Process}
\ac{SAM2} comes with four different \enquote{checkpoints}, \ie, model sizes.
For \AppName, we chose to focus on the \texttt{large} checkpoint, which has the most tunable parameters, as it yielded the most promising results in initial experiments.
We performed fine-tuning using a single Nvidia H100 machine-learning accelerator for each run.
For all our experiments, we set a fixed batch size of $12$, as this is the maximum that fits in the \SI{80}{GB} of memory on our Nvidia H100 SXM cards.
Our fine-tuning scripts utilized PyTorch v2.7.1 and CUDA 13.0, and fine-tuning on \SI{90}{\percent} of our dataset took around \SI{69}{\hour} to complete.

The fine-tuning process is as follows: 
First, we retrieve segmentation masks from the current model.
To this end, we feed the input images of each batch to the image encoder.
For each of these images, we then get embeddings of a point prompt from the prompt encoder.
Here, we sample a random point $(x,y)$ from the ground-truth foreground (\ie, the point lies on a metal line) and use it as a positive prompt to the prompt encoder.
Next, we feed the image and prompt embeddings to the mask decoder, which produces low-resolution masks. 
These masks are then postprocessed by internal \ac{SAM2} functions to obtain three prediction masks for each image in the batch, as we enabled multi-mask output for better segmentation results~\cite{DBLP:conf/iclr/RaviGHHR0KRRGMP25}.
Finally, we compute the loss between the ground truth and the segmentation mask, then perform backward propagation and optimize the targeted model components using AdamW~\cite{DBLP:conf/iclr/LoshchilovH19}.

\subsubsection{Data Augmentation}
\label{mseg::subsubsection::data_aug}
We leverage data augmentation during fine-tuning to make our fine-tuned \ac{SAM2} model more robust to small perturbations in input images and to expand the training dataset.
To this end, we use Torchvision from PyTorch to apply transforms to the training data before feeding it to the fine-tuning algorithms. 
Below, we list the individual transforms that we deploy in the order in which they are applied to the input data:
\begin{enumerate}
    \item \texttt{RandomResizedCrop}: $\texttt{size} = 1024$, $\texttt{scale} = [0.99-0.5 \cdot \texttt{intensity},0.99]$
    \item \texttt{GaussianBlur}: $\texttt{kernel\_size}=(5, 9)$, $\texttt{sigma}=5.0 \cdot \texttt{intensity}$
    \item \texttt{ColorJitter}: $\texttt{brightness}=0.75 \cdot \texttt{intensity}$, $\texttt{contrast}=0.5 \cdot \texttt{intensity}$
    \item \texttt{RandomVerticalFlip}: $\texttt{probability}=0.5 \cdot \texttt{intensity}$
    \item \texttt{RandomHorizontalFlip}: $\texttt{probability}=0.5 \cdot \texttt{intensity}$
    \item \texttt{GaussianNoise}: $\texttt{mean}=0.0$, $\texttt{sigma}=0.5 \cdot \texttt{intensity}$, $\texttt{clip}=\text{True}$
\end{enumerate}

The specific parameters were manually selected by domain experts to serve as the upper bound for data augmentation.
The experts selected these parameters so that, for an intensity value of $1$, the resulting images could still (but barely) be interpreted by a human analyst.
All transforms share a common data augmentation intensity value that can be set to any value in $(0, 1]$ and is determined by our hyperparameter optimization, as shown in \Cref{mseg::subsubsection::hyperparam}.
During each iteration, we randomly apply all data augmentation transforms to the input image or none at all.
The probability of applying data augmentation to an image is also determined through hyperparameter optimization.

\subsubsection{Loss Functions}
\label{mseg::subsubsection::loss_functions}
For our pixel-based loss function $L_\text{pixel}$, we choose a combination of the \ac{BCE} loss $L_\text{BCE}$~\cite{DBLP:conf/icml/MaoM023a} and the Dice loss $L_\text{Dice}$~\cite{DBLP:conf/miccai/SudreLVOC17} based on a blending parameter $\alpha \in (0,1)$ that is determined by our hyperparameter search:
$$L_\text{pixel} = \alpha \cdot L_\text{BCE} + (1-\alpha) \cdot L_\text{Dice}.$$
\ac{SAMIC}~\cite{DBLP:conf/tencon/NgTCG24} additionally employs the \ac{IoU} loss~\cite{DBLP:conf/cvpr/RezatofighiTGS019}, but our segmentation accuracy did not improve when incorporating the \ac{IoU} into $L_\text{pixel}$.

We want to promote correct connectivity and penalize shorts and open circuits during training to reduce the \ac{ESD} error rate.
Since the \ac{ESD} error computation itself is not differentiable, it cannot be used as a loss function right away.
Instead, we need to find a differentiable loss term that behaves similarly to the \ac{ESD} error.
To this end, we supplement our pixel-based loss with a topological loss.
After careful consideration and consultation with experts, we selected the Betti matching loss~\cite{DBLP:conf/icml/StuckiPSMB23}, $L_\text{Betti}$, for this purpose.
We also explored two other topology-preserving loss functions, namely TopoLoss by Hu \etal~\cite{DBLP:conf/nips/HuLSC19} and Topograph by Lux \etal~\cite{DBLP:conf/iclr/LuxBWSRBP25}, but decided against them for runtime and effectiveness reasons.
Generally, such loss functions focus on correct connectivity, but still need to be combined with pixel-based losses to enforce local accuracy.
In the following, we only consider the application of Betti matching to images.

The Betti matching loss is based on concepts from persistent homology, which examines how topological features emerge and disappear as a filtration parameter $\epsilon$ varies.
In particular, it tracks when topological features, such as connected components, cycles, or holes, are born and die along the filtration.
This filtration can, for example, be induced by thresholding a scalar-valued function such as the model’s prediction likelihoods.
The number of detected topological structures (\eg, corresponding to metal lines in our images) then depends on the chosen threshold value.
For each topological feature, a pair $(b, d)$ is stored that records its birth time $b$ (\ie, the value of $\epsilon$ at which the feature first appears) and its death time $d$ (\ie, the value of $\epsilon$ at which it merges into an older feature or becomes topologically trivial).
Tracking many such features simultaneously produces a so-called persistence barcode, where each bar represents the birth and death times of a single feature.
Betti numbers $\beta_k(\epsilon)$ summarize this information by counting the number of $k$-dimensional topological features that are alive at a given value of $\epsilon$.
For images, we only consider $k \in \{0,1\}$, where $\beta_0(\epsilon)$ counts connected components and $\beta_1(\epsilon)$ counts loops or holes.
The Betti matching loss goes one step further by spatially matching topological features from one image (e.g., the ground truth $G$) to corresponding features in another image (e.g., the prediction mask $L$).
To this end, persistence barcodes are computed for both images.
For the loss computation, an optimal matching between the two persistence barcodes is determined, and the resulting matching error is calculated.
This error is then used to construct a differentiable, efficient, and topology-aware loss function.

The Betti matching loss $L_\text{Betti}$ is already implemented as a Python package\footnote{\url{https://pypi.org/project/topolosses/}} that is based on the work of Stucki \etal~\cite{DBLP:conf/icml/StuckiPSMB23,DBLP:journals/corr/abs-2407-04683} and Berger \etal~\cite{DBLP:conf/miccai/BergerLSBSBRBP24}, making it straightforward to use.
It comes with a few tweakable parameters relevant to our hyperparameter search.
Firstly, the \texttt{filtration\_type} defines how the comparison image $C$ is computed as $C \leftarrow \min(G, L)$, \ie, the element-wise minimum of ground truth $G$ and prediction mask $L$, which in turn governs the Betti matching process. 
It can take the values \texttt{superlevel} (features appear as input values decrease), \texttt{sublevel} (features appear as input values increase), and \texttt{bothlevels} (applies both filtration types and combines results).
Secondly, the parameter \texttt{barcode\_length\_threshold} governs the robustness of the Betti matching loss against short-lived topological features that may arise from unclean or noisy prediction masks. 
Lastly, \texttt{push\_unmatched\_to\_1\_0} determines whether unmatched birth points are pushed toward 1 while unmatched death points are pushed toward 0 (\texttt{True}) or whether both points are just pushed together (\texttt{False}).

In the end, we construct our final segmentation loss $L_\text{seg}$ used for fine-tuning as 
$$L_\text{seg} = L_\text{pixel} + \lambda \cdot L_\text{Betti}$$
with $\lambda \in [0, 1]$ being a blending parameter determined by the hyperparameter search.

\subsection{Hyperparameter Optimization}
\label{mseg::subsubsection::hyperparam}
To determine the best-suited hyperparameters for fine-tuning \ac{SAM2}, we conducted a hyperparameter search using the hyperband pruner from the Optuna framework~\cite{DBLP:conf/kdd/AkibaSYOK19}.
To this end, we initiated dozens of fine-tuning runs, each with different sets of hyperparameters, to determine the best-fitting parameters for our setting.
All of these fine-tuning runs were executed on a cluster of eight Nvidia H100 machine-learning accelerators, which comprises a mix of SXM cards with \SI{80}{GB} and NVL cards with \SI{94}{GB} of memory.
Still, only a single H100 was used for each fine-tuning run; the cluster was used only to run multiple sessions in parallel.
The hyperband pruner attempts to minimize an objective function, specifically the relative number of \ac{ESD} errors in relation to the total number of metal lines, as computed on our validation dataset from \acp{IC} 1 to 7.
In total, the hyperparameter search ran for seven days.

\subsubsection{Setup}
Due to runtime limitations, we opted for a two-step hyperparameter optimization: 
(i)~First, we searched for the best parameters related to data augmentation and pixel-based loss functions by repeatedly fine-tuning \ac{SAM2} on our training dataset from \acp{IC} 1 to 7.
To this end, we set the maximum number of epochs per hyperparameter optimization run to 25 and allowed pruning only after 5 epochs.
During this first step, we set $\lambda=0$ and hence $L_\text{seg} = L_\text{pixel}$.
This is in line with the recommendations of Berger \etal to add Betti matching to the loss only after pixel-based losses have stabilized to achieve performance gains.
To identify the best parameter set, we evaluated each model after every epoch using our validation dataset from \acp{IC} 1 to 7.
We then selected the best-performing parameter set by analyzing the performance of all fine-tuned models at every epoch.
Using these selected parameters, we fine-tuned the off-the-shelf \ac{SAM2} for 50 epochs.
Here, we found that this fine-tuned model performs best after 35 training epochs; hence, we selected this version as a starting point for the second step. 
(ii)~Finally, we enabled the Betti matching loss for continued fine-tuning of this selected model and started a second hyperparameter optimization to determine suitable hyperparameters for this topological loss function.
The second hyperparameter optimization ran for 15 additional epochs, bringing the total to up to 50 fine-tuning epochs per trial across both steps.

\subsubsection{Results}
For the first hyperparameter optimization step (i), we ran 56 Optuna trials, each yielding a fine-tuned model trained for up to 25 epochs.
In total, 49 trials were pruned by Optuna before completing all epochs. 
For the best trial, the objective function, \ie, the percentage of \ac{ESD} errors in relation to the total number of metal lines, reached its minimum at epoch 22 with a value of \SI{0.659}{\percent}.
Based on our evaluation, we determined the probability of data augmentation being applied to be \SI{38.5}{\percent} and the data augmentation intensity to be $0.61$.
For the pixel-based loss function $L_\text{pixel}$ and the AdamW optimizer, we fix the learning rate at $1.111 \cdot 10^{-5}$, the weight decay at $2.059 \cdot 10^{-6}$, and a blending parameter value of $\alpha=0.6$.
Furthermore, we observed that the learning rate has the greatest impact on segmentation performance, with values above $10^{-3}$ often leading to local minima, resulting in the fine-tuned model producing only completely black segmentation masks.

In the second step (ii), we conducted a total of 75 Optuna trials over 5 days, with 58 trials being pruned before completion.
For the best trial, we report $\lambda = 0.375$ as the blending parameter for $L_\text{seg}$.
Furthermore, we determined the optimal parameter set for the Betti matching loss in our application.
To this end, we choose \texttt{sublevel} as the \texttt{filtration\_type}, $0.345$ as the \texttt{barcode\_length\_threshold}, and enabled pushing unmatched birth points to 1 and death points to 0 (\texttt{push\_unmatched\_to\_1\_0}).

\subsection{Segmenting Images}
\label{mseg::subsection::inference}
Besides the input image, \AppName also requires at least one point prompt pinpointing the structure to segment.
To generate an effective point prompt, we need to ensure that the positive prompt we provide is actually located on one of the metal lines in the input image.
To this end, we use lightweight classical image analysis on the input images fed to both models \BoxOne and \BoxTwo to identify parts that are highly likely to be foreground, \ie, belong to a metal line.
We do not aim for a fully accurate segmentation at this point; we just need to identify some foreground areas, but not all of them.
The binary mask is generated by aggressively thresholding the brightest pixels in the input image.

In addition, we denoise, remove small objects, and apply light morphological operations to eliminate isolated pixels, fill tiny gaps, and smooth object boundaries.
We then place five random point prompts within the identified foreground areas and feed them to the fine-tuned model along with the input image.
In our experiments, we observed that providing more point prompts does not improve segmentation results.

With a batch size of 12, segmentation of a single patch on one H100 takes \SI{0.92}{\second} and uses up to \SI{8471}{\mebi\byte} of \acs{GPU} memory.
Segmenting original-size images takes a few milliseconds longer, likely due to internal resizing.
Internally, \AppName produces one segmentation mask for each point prompt, along with scores rating their quality.
We always return the mask with the highest score as the final segmentation mask.
We observe that segmentation time scales linearly with the number of point prompts. 
For example, reducing this number from five to two results in a segmentation time of \SI{0.4}{\second} per patch.

\section{Evaluation}
\label{mseg::section::eval}
First, we evaluate \AppName's in-distribution performance using previously unseen \ac{SEM} images from \acp{IC}~1 to~7 in \Cref{mseg::subsection::segperf}.
Next, we investigate \AppName's out-of-distribution performance on metal layer images from unseen \acp{IC} in \Cref{mseg::subsection::genperf}.
We then analyze failure cases in \Cref{mseg::subsection::outliers}, test different dataset splits with reduced numbers of \acp{IC} for fine-tuning to better understand how \AppName generalizes to unseen \acp{IC} in \Cref{mseg::subsection::generalization_splits}, and present its segmentation accuracy when trained on our full dataset of 14 \acp{IC} in \Cref{mseg::subsection::full_data}.

\subsection{In-Distribution Performance on Seen ICs}
\label{mseg::subsection::segperf}
In this section, we evaluate the in-distribution performance of \AppName and compare it to previous work and established segmentation approaches.
To this end, we used the training datasets for \acp{IC} 1 to 7 established in \Cref{mseg::subsection::dataset} for all considered techniques.
To compare \AppName with other techniques, we fine-tuned SAMIC using the code of Ng. \etal~\cite{DBLP:conf/tencon/NgTCG24}.
We also applied the unsupervised method proposed by Rothaug \etal~\cite{DBLP:conf/ashes/RothaugKABP0P23} to our dataset and trained U-Net, DeepLabV3, and FCN using their code.
We then evaluated each model on our test dataset from \acp{IC} 1 to 7 and measured the segmentation accuracy against the ground truth. 
Hence, in this experiment, we assess how well \AppName performs on metal-layer images structurally similar to those used for training.
While we evaluated most approaches on original-size images featuring a total of \numprint{36413} metal lines, the approach by Rothaug \etal is evaluated on patches due to its design.
These patches feature a total of \numprint{89063} metal lines, as some metal lines from the original-size images are present on multiple patches, thereby explaining the higher absolute number of \ac{ESD} errors compared to other approaches.
Refer to \Cref{mseg::table::segmentation} for the results.

\begin{table}[htb]
    \centering
    \footnotesize
    \begin{tabularx}{\linewidth}{X|C{.8cm}C{.75cm}C{.75cm}C{.8cm}C{.8cm}C{.75cm}|C{.8cm}C{.8cm}C{.8cm}}
        \toprule
        & \multicolumn{6}{c|}{\textbf{ESD} $\downarrow$} & \multicolumn{3}{c}{\textbf{pixel} $\uparrow$}\\
        & total & \% & opens & shorts & FPs & FNs & PA & Dice & IoU \\
        \midrule
        \textbf{\AppName} & \underline{263} & \underline{0.72} & 27 & \underline{146} & 59 & 31 & \underline{0.972} & 0.960 & 0.925 \\
        - only pixel loss & 323 & 0.88 & 29 & 211 & \underline{54} & 29 & \underline{0.972} & \underline{0.961} & \underline{0.926} \\
        - single model & 289 & 0.79 & \underline{22} & 185 & 59 & \underline{23} & \underline{0.972} & \underline{0.961} & 0.925 \\
        - only \BoxOne & \numprint{1914} & 5.25 & 223 & \numprint{1519} & 96 & 76 & 0.950 & 0.926 & 0.865 \\
        - only \BoxTwo & 286 & 0.78 & 27 & \underline{146} & 82 & 31 & 0.954 & 0.940 & 0.900 \\
        \midrule
        SAMIC & \numprint{2178} & 5.98 & 412 & \numprint{1083} & 649 & 34 & 0.961 & 0.944 & 0.899 \\
        Rothaug \etal & \numprint{13086} & 14.69 & \numprint{5424} &\numprint{1144} & \numprint{5972} & 546 & 0.939 & 0.901 & 0.831\\
        U-Net & \numprint{1618} & 4.44 & 78 & 764 & 735 & 41 & 0.966 & 0.949 & 0.905 \\
        DeepLabV3 & \numprint{2187} & 6.01 & 236 & \numprint{1705} & 212 & 34 & 0.961 & 0.946 & 0.901 \\
        FCN & \numprint{2067} & 5.68 & 277 & \numprint{1443} & 316 & 31 & 0.963 & 0.950 & 0.907 \\
        \bottomrule
    \end{tabularx}
    \caption{\textbf{In-distribution} performance of \AppName compared to other approaches. 
    All models were trained and evaluated on the same training dataset from \acp{IC} 1--7.
    We report the \ac{ESD} error rate in two forms: as the absolute number of \ac{ESD} errors observed for our test set, and as the relative number of \ac{ESD} errors per metal line, expressed as a percentage. 
    Furthermore, we present commonly used metrics for pixel-level segmentation accuracy.}
    \label{mseg::table::segmentation}
\end{table}

As we are more interested in electrical correctness than in pixel-accuracy, we follow Trindade \etal~\cite{DBLP:conf/ccece/TrindadeUSP18} and adopt their \ac{ESD} metric for evaluation.
We leverage the implementation of Rothaug \etal~\cite{DBLP:conf/ashes/RothaugKABP0P23,DBLP:journals/jce/RothaugCKABPBP25} to count the number of opens, shots, \acp{FP}, and \acp{FN} in the segmented images compared to the ground truth.
To this end, we report the absolute number of \ac{ESD} errors observed in our test set, the relative number of \ac{ESD} errors per metal line, and the counts of opens, shorts, \acp{FP}, and \acp{FN}.
All evaluations are conducted on the original-size images.
For \AppName, we also provide evaluation results for when only the pixel-based loss is used, as well as for each of the two models, \BoxOne (operating on the original-size images) and \BoxTwo (operating on $512 \times 512$-pixel patches).
Furthermore, we analyze the performance when fine-tuning a single model on both original-sized images and smaller patches.
We are not yet evaluating \AppName's out-of-distribution performance; hence, we test on the \acp{IC} we also used for training.

\paragraph{Results.}
From \Cref{mseg::table::segmentation}, we observe that \AppName is on-par with SAMIC, U-Net, DeepLabV3, and FCN for in-distribution segmentation in terms of the pixel-based metrics, but produces significantly fewer \ac{ESD} errors.
In particular, \AppName drastically reduces the number of opens, shorts, and \acp{FP} in the segmentation by about a factor of 6 relative to U-Net, which is the next-best approach.
Furthermore, we clearly see the benefit of our multi-scale approach, as models~\BoxOne and \BoxTwo individually do not achieve comparable results.
For in-distribution analysis, we do not observe a significant impact of using a single model that combines models \BoxOne and \BoxTwo.
The \ac{ESD} error rate increases marginally, which could still be tolerated in exchange for a more straightforward fine-tuning process.
Finally, we can see a clear benefit of using our Betti matching loss function over pixel-based losses alone.

\subsection{Out-of-Distribution Performance on Unseen ICs}
\label{mseg::subsection::genperf}
To assess how well \AppName performs on previously unseen \acp{IC}, we utilize the models previously fine-tuned on the training dataset from \acp{IC} 1 to 7 for the in-distribution evaluation in \Cref{mseg::subsection::segperf}.
This time, however, we test its out-of-distribution performance on all available images from \acp{IC} 8 to 14.
As these \acp{IC} are not used for training, we can safely use all their metal-layer images for testing.
For comparison, we again evaluate SAMIC, the unsupervised approach by Rothaug \etal~\cite{DBLP:conf/ashes/RothaugKABP0P23}, U-Net, DeepLabV3, and FCN on our dataset.
The original-size images feature a total of \numprint{15153} metal lines.
As the approach from Rothaug \etal gains its generalization capabilities through unsupervised training, we trained it on \SI{70}{\percent} of the image patches from \acp{IC} 8 to 14, used \SI{10}{\percent} of that dataset for validation, and the remaining \SI{20}{\percent} for evaluation.
Hence, in this case, the evaluation patches contained only \numprint{10244} metal lines.
Following the same evaluation procedure as described in \Cref{mseg::subsection::segperf}, we report results from these experiments in \Cref{mseg::table::generalization}. 

\begin{table}[htb]
    \centering
    \footnotesize
    \begin{tabularx}{\linewidth}{X|C{.75cm}C{.75cm}C{.75cm}C{.8cm}C{.8cm}C{.75cm}|C{.8cm}C{.8cm}C{.8cm}}
        \toprule
        & \multicolumn{6}{c|}{\textbf{ESD} $\downarrow$} & \multicolumn{3}{c}{\textbf{pixel} $\uparrow$}\\
        & total & \% & opens & shorts & FPs & FNs & PA & Dice & IoU \\
        \midrule
        \textbf{\AppName} & 838 & 5.53 & 531 & \underline{170} & 104 & \underline{33} & \underline{0.959} & 0.939 & 0.888 \\
        - only pixel loss & \underline{789} & \underline{5.21} & 351 & 234 & 164 & 40 & \underline{0.959} & \underline{0.944} & \underline{0.896} \\
        - single model & 908 & 5.99 & 410 & 235 & 204 & 59 & 0.958 & 0.937 & 0.886 \\
        - only \BoxOne & \numprint{1607} & 10.60 & \underline{340} & \numprint{1112} & \underline{72} & 83 & 0.936 & 0.920 & 0.854 \\
        - only \BoxTwo & \numprint{1443} & 9.52 & 649 & 468 & 280 & 46 & 0.934 & 0.911 & 0.852 \\
        \midrule
        SAMIC & \numprint{4116} & 27.16 & \numprint{2079} & 363 & \numprint{1608} & 66 & 0.934 & 0.935 & 0.831\\
        Rothaug \etal & \numprint{2280} & 22.25 & \numprint{1640} & 291 & 245 & 104 & 0.911 & 0.868 & 0.788\\
        U-Net & \numprint{7704} & 44.90 & \numprint{1570} & 603 & \numprint{4435} & 196 & 0.907 & 0.830 & 0.753\\
        DeepLabV3 & \numprint{3753} & 24.77 & \numprint{2462} & 154 & 437 & 700 & 0.904 & 0.787 & 0.714\\
        FCN & \numprint{5815} & 38.38 & \numprint{3158} & 303 & \numprint{2006} & 348 & 0.910 & 0.826 & 0.751\\
        \bottomrule
    \end{tabularx}
    \caption{\textbf{Out-of-distribution} performance of \AppName compared to other approaches. 
    All models were trained on the same training dataset from \acp{IC} 1--7 and evaluated on \acp{IC} 8--14.
    We report the \ac{ESD} error rate in two forms: as the absolute number of \ac{ESD} errors observed for our test set, and as the relative number of \ac{ESD} errors per metal line, expressed as a percentage. 
    Furthermore, we present commonly used metrics for pixel-level segmentation accuracy.}
    \label{mseg::table::generalization}
\end{table}

\paragraph{Results.} 
Examining the out-of-distribution results in \Cref{mseg::table::generalization}, we can see that \AppName now fully plays to its strengths.
While existing work exhibits, at best, an \ac{ESD} error rate of \SI{22.25}{\percent} for unseen \acp{IC}, meaning that about every fourth segmented metal line is incorrect, \AppName achieves an error rate of just \SI{5.53}{\percent}.
Again, we see that using either model~\BoxOne or model~\BoxTwo in isolation would not yield satisfactory results, although both still perform better than other approaches from the literature.
Now, we can also observe degraded performance if we combine model~\BoxOne and \BoxTwo into a single one.
In that case, the \ac{ESD} error rate for generalization would increase by about 0.5 percentage points.
Finally, using only the pixel-based loss yields a slight improvement in segmentation accuracy.
Looking more closely, \AppName with Betti matching loss results in more open circuits, but fewer shorts and \acp{FP} than the model fine-tuned without any topological loss.

\subsection{Analysis of Failure Cases}
\label{mseg::subsection::outliers}
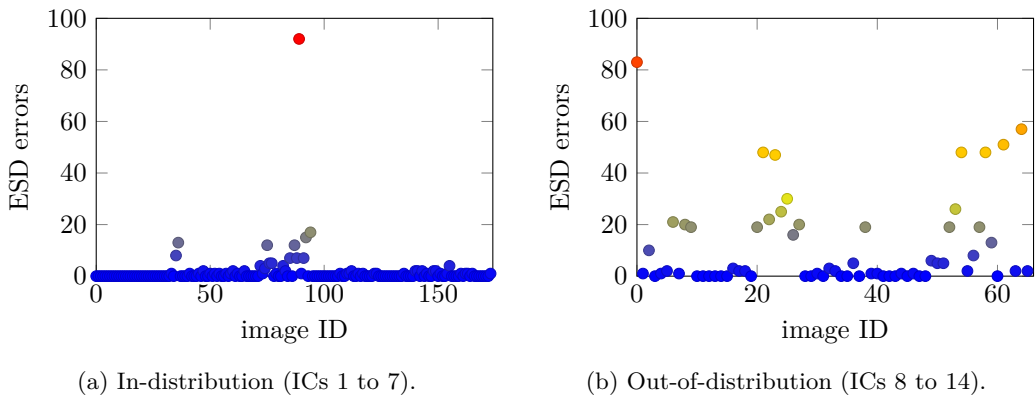
\begin{figure}[htb]
  \centering
  \begin{subfigure}{0.48\textwidth}
    \centering
    \begin{tikzpicture}
        \begin{axis}[
            width=6.8cm,
            height=4cm,
            xlabel={image ID},
            ylabel={ESD errors},
            xlabel style={yshift=2pt},
            ylabel style={yshift=-6pt},
            ymin=0, ymax=100,
            xmin=0, xmax=174,
            only marks,
        ]
            \addplot[
                scatter,
                mark=*,
            ]
            table[
                y index=0,
                x expr=\coordindex
            ] {sections/data/image_seg_res.dat};
        \end{axis}
    \end{tikzpicture}
    \caption{In-distribution (\acp{IC} 1 to 7).}
    \label{mseg::subfigure::image_seg_esd}
  \end{subfigure}\hfill
  \begin{subfigure}{0.48\textwidth}
    \centering
    \begin{tikzpicture}
        \begin{axis}[
            width=6.8cm,
            height=4cm,
            xlabel={image ID},
            ylabel={ESD errors},
            xlabel style={yshift=2pt},
            ylabel style={yshift=-6pt},
            ymin=0, ymax=100,
            xmin=0, xmax=66,
            only marks,
        ]
            \addplot[
                scatter,
                mark=*,
            ]
            table[
                y index=0,
                x expr=\coordindex
            ] {sections/data/image_gen_res.dat};
        \end{axis}
    \end{tikzpicture}
    \caption{Out-of-distribution (\acp{IC} 8 to 14).}
    \label{mseg::subfigure::image_gen_esd}
  \end{subfigure}

  \caption{Distribution of the absolute number of \ac{ESD} errors across all original-size images.}
  \label{mseg::figure::image_esd}
\end{figure}

We now take a closer look at the distribution of \ac{ESD} errors across the test images used in our in-distribution (\acp{IC} 1 to 7) and out-of-distribution (\acp{IC} 8 to 14) evaluation.
For in-distribution, we observe only one outlier image, exhibiting 92 \ac{ESD} errors, as shown in \Cref{mseg::subfigure::image_seg_esd}.
Although we observe more outliers in our out-of-distribution evaluation, we can still report decent results for most images, as shown in \Cref{mseg::subfigure::image_gen_esd}.
Here, we count 7 images with more than 40 \ac{ESD} errors and another 5 with more than 20 errors.

\begin{figure}[htbp]
  \centering
  \begin{subfigure}{0.3\linewidth}
    \centering
    \includegraphics[width=\linewidth]{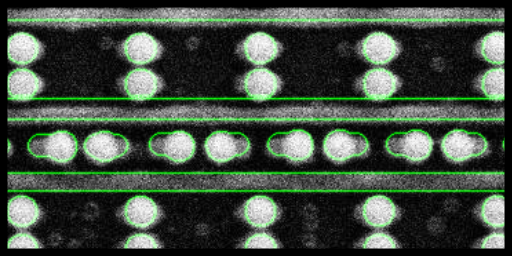}
    \caption{Misalignment of ground truth (green) and image.}
    \label{mseg::subfigure::bad_image_1}
  \end{subfigure}
  \hfill
  \begin{subfigure}{0.3\linewidth}
    \centering
    \includegraphics[width=\linewidth]{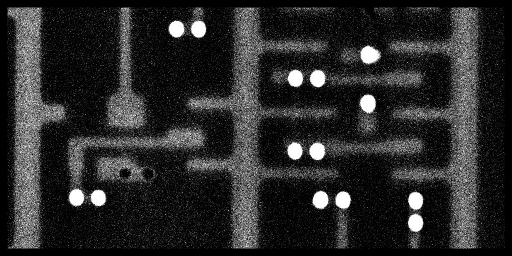}
    \caption{Dark shades around vias and missing vias.}
    \label{mseg::subfigure::bad_image_2}
  \end{subfigure}
  \hfill
  \begin{subfigure}{0.3\linewidth}
    \centering
    \includegraphics[width=\linewidth]{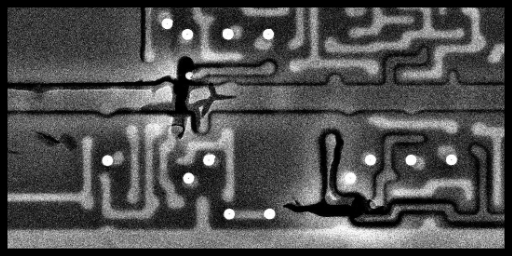}
    \caption{Delayering defects, only outlines of metal lines visible.}
    \label{mseg::subfigure::bad_image_3}
  \end{subfigure}

  \caption{Faulty images with artifacts or defects from sample preparation or imaging.}
  \label{mseg::figure::bad_images}
\end{figure}

We now examine some of these failure cases in more detail to identify the reasons behind the observed results. 
Across our sample, we mostly observe three sources of errors: 
\begin{enumerate}[label=(\roman*)]
    \item For the outlier in \Cref{mseg::subfigure::image_seg_esd}, we observe a slight misalignment between the actual metal lines on the \ac{SEM} images and the ground truth mask, as shown in \Cref{mseg::subfigure::bad_image_1}.
    Here, the \ac{ESD} metric simply counts invalid overlaps between a metal line in the ground truth and vias in the predicted segmentation mask as shorts.
    \item The first outlier in \Cref{mseg::subfigure::image_gen_esd} results from two via-related issues.
    Existing vias in that image cast a black shade around them, see \Cref{mseg::subfigure::bad_image_2}.
    They partially obscure the underlying metal line and mislead \AppName into predicting background around the via.
    We observed the same issue for other outliers in the out-of-distribution test, which can be explained by the absence of any images with similar characteristics in our training set.
    For \Cref{mseg::subfigure::bad_image_2}, we also observe that some vias are missing altogether, as they were accidentally removed during sample preparation.
    This results in narrow metal lines around the missing via that \AppName does not properly segment.
    \item Some failures around image 54 in \Cref{mseg::subfigure::image_gen_esd} are due to delayering defects, see \Cref{mseg::subfigure::bad_image_3}.
    Some metal lines were incorrectly removed during sample preparation, leaving only their outlines.
    These failures increase the \ac{ESD} error rate across all affected images.
\end{enumerate}

Of course, not all \ac{ESD} errors can be attributed to such sample preparation and image quality issues. 
However, our out-of-distribution test set from \acp{IC} 8 to 14 is of lower quality than \acp{IC} 1 to 7, which helps to explain the more frequent significant outliers in \Cref{mseg::subfigure::image_gen_esd}.

\subsection{More ICs Result in Better Generalization}
\label{mseg::subsection::generalization_splits}
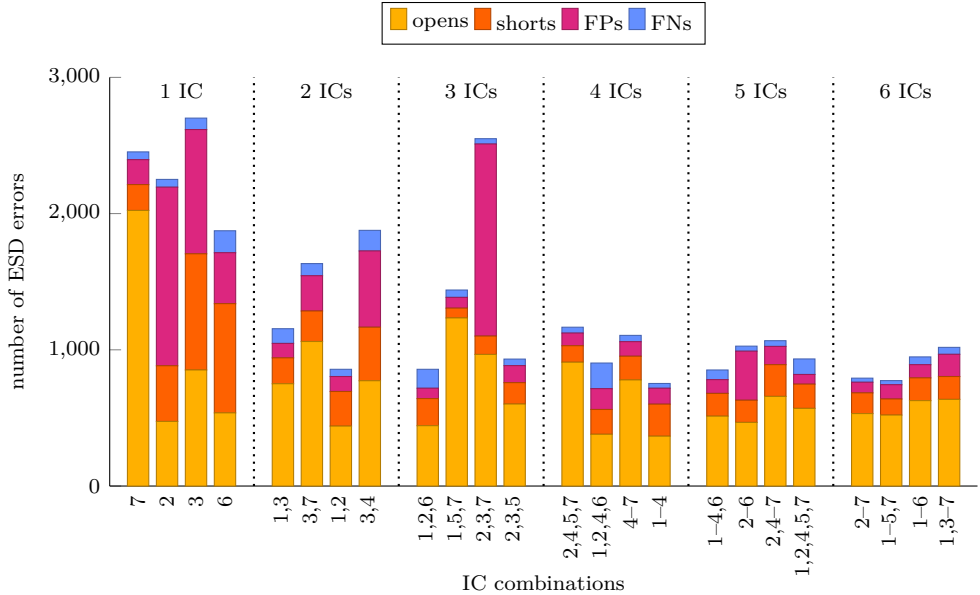
\begin{figure}
    \centering
    \begin{tikzpicture}
        \begin{axis}[
            ybar stacked,
            width=13cm,
            height=6cm,
            ymin=0,
            ymax=3000,
            bar width=8pt,
            enlarge x limits=0.035,
            axis x line*=bottom,
            axis y line*=left,
            xlabel={IC combinations},
            ylabel={number of ESD errors},
            xtick=data,
            xticklabels={
                7,2,3,6,
                {1,3},{3,7},{1,2},{3,4},
                {1,2,6},{1,5,7},{2,3,7},{2,3,5},
                {2,4,5,7},{1,2,4,6},{4--7},{1--4},
                {1--4,6},{2--6},{2,4--7},{1,2,4,5,7},
                {2--7},{1--5,7},{1--6},{1,3--7}
            },
            xticklabel style={rotate=90, anchor=east, font=\footnotesize},
            yticklabel style={font=\footnotesize},
            xlabel style={font=\footnotesize,yshift=-15pt},
            ylabel style={font=\footnotesize},
            legend style={
                at={(0.98,0.88)},
                anchor=north east,
                legend columns=1,
                font=\footnotesize
            },
        ]

            \addplot [fill=ibm_yellow,draw=ibm_yellow!70!black] coordinates {
                (0,2023) (0.5,474) (1,852) (1.5,537)
                (2.5,751) (3,1061) (3.5,440) (4,773)
                (5,443) (5.5,1234) (6,966) (6.5,602)
                (7.5,909) (8,380) (8.5,779) (9,366)
                (10,513) (10.5,467) (11,658) (11.5,570)
                (12.5,532) (13,521) (13.5,627) (14,636)
            };
            
            \addplot [fill=ibm_orange,draw=ibm_orange!70!black] coordinates {
                (0,190) (0.5,409) (1,853) (1.5,802)
                (2.5,190) (3,224) (3.5,254) (4,393)
                (5,199) (5.5,72) (6,135) (6.5,157)
                (7.5,121) (8,182) (8.5,174) (9,235)
                (10,167) (10.5,163) (11,233) (11.5,178)
                (12.5,152) (13,118) (13.5,167) (14,168)
            };
            
            \addplot [fill=ibm_magenta,draw=ibm_magenta!70!black] coordinates {
                (0,183) (0.5,1311) (1,912) (1.5,374)
                (2.5,106) (3,259) (3.5,110) (4,560)
                (5,76) (5.5,79) (6,1410) (6.5,125)
                (7.5,93) (8,153) (8.5,107) (9,117)
                (10,101) (10.5,360) (11,134) (11.5,71)
                (12.5,78) (13,105) (13.5,97) (14,163)
            };
            
            \addplot [fill=ibm_blue,draw=ibm_blue!70!black] coordinates {
                (0,57) (0.5,57) (1,84) (1.5,161)
                (2.5,108) (3,89) (3.5,53) (4,151)
                (5,139) (5.5,54) (6,39) (6.5,48)
                (7.5,43) (8,188) (8.5,46) (9,35)
                (10,71) (10.5,37) (11,42) (11.5,114)
                (12.5,30) (13,30) (13.5,57) (14,51)
            };
            
            \legend{opens, shorts, FPs, FNs}
            
            \draw[dotted, thick]
                (axis cs:2,0) -- (axis cs:2,3000)
                (axis cs:4.5,0) -- (axis cs:4.5,3000)
                (axis cs:7,0) -- (axis cs:7,3000)
                (axis cs:9.5,0) -- (axis cs:9.5,3000)
                (axis cs:12,0) -- (axis cs:12,3000);
            
            \node[font=\footnotesize]
                at (axis cs:0.75,2900) {1 IC};
            \node[font=\footnotesize]
                at (axis cs:3.25,2900) {2 ICs};
            \node[font=\footnotesize]
                at (axis cs:5.75,2900) {3 ICs};
            \node[font=\footnotesize]
                at (axis cs:8.25,2900) {4 ICs};
            \node[font=\footnotesize]
                at (axis cs:10.75,2900) {5 ICs};
            \node[font=\footnotesize]
                at (axis cs:13.25,2900) {6 ICs};
            
        \end{axis}
    \end{tikzpicture}
    \caption{Different training data splits used for fine-tuning to investigate whether more diversity in \acp{IC} during fine-tuning affects the segmentation accuracy on unseen \acp{IC}.}
    \label{mseg::figure::gen_splits}
\end{figure}

We now investigate how the number of \acp{IC} seen during fine-tuning affects the out-of-distribution segmentation performance on unseen \acp{IC}.
To this end, we fine-tune \ac{SAM2} on different subsets of \acp{IC} 1 to 7, ranging from a single \ac{IC} to six different \acp{IC}.
For each number of \acp{IC}, we fine-tune on four different randomly chosen \ac{IC} combinations and then always infer on the same out-of-distribution test set from \acp{IC} 8 to 14.
As becomes evident from the resulting numbers of \ac{ESD} errors in \Cref{mseg::figure::gen_splits}, the more \acp{IC} are used for fine-tuning, the better the out-of-distribution performance.
Intuitively, this was expected.
However, our results also indicate that the \ac{ESD} error rate has not yet saturated, and this downward trend is likely to continue when using even more \acp{IC} for fine-tuning.

\subsection{Fine-Tuning on All 14 ICs}
\label{mseg::subsection::full_data}
After carefully evaluating the in-distribution and out-of-distribution capabilities of \AppName when fine-tuning on seven \acp{IC}, we fine-tuned on \SI{90}{\percent} of all annotated images from all 14 \acp{IC}.
As our findings from \Cref{mseg::subsection::generalization_splits} imply that \AppName's generalization accuracy has not yet saturated, we expect this final model to achieve even better segmentation accuracy.
To test its in-distribution performance, we allocated \SI{10}{\percent} of all images from all \acp{IC} for testing.
When evaluating on this \SI{10}{\percent} test set, we obtain a \ac{PA} of 0.971 (Dice: 0.960, IoU: 0.924) and an \ac{ESD} error rate of \SI{0.62}{\percent} (22 shorts, 50 opens, 41 \acp{FP}, and 6 \acp{FN} across $19,088$ metal lines). 
Hence, we observe a slight improvement over the \SI{0.72}{\percent} in-distribution \ac{ESD} error rate reported for seven \acp{IC}.
Given that we now use all our \acp{IC} for fine-tuning, we cannot re-evaluate out-of-distribution performance, as the model has already seen all \acp{IC} in our dataset.
However, as we publish this final model as part of our work, interested readers can test it on their own \ac{IC} images without further fine-tuning.
\section{Discussion and Conclusion}
\label{mseg::section::conclusion}
We now discuss limitations of our approach in \Cref{mseg::subsection::limitations}, potential avenues for future research in \Cref{mseg::subsection::future_work}, and present our concluding remarks in \Cref{mseg::subsection::conclude}.

\subsection{Limitations}
\label{mseg::subsection::limitations}
One major limitation is that our ground truth is based on manually annotated \ac{SEM} image data.
In rare cases, it was even difficult, if not impossible, for the human creating the ground truth to decide whether metal lines were connected.
Hence, some remaining errors can be attributed to uncertainties in the ground truth.
The only viable solution for this problem would be to use the design files of the analyzed \acp{IC} as a ground truth, which were not available to us for any of the targeted third-party \acp{IC}.

We agree that an \ac{ESD} error rate of \SI{5.53}{\percent} on unseen \acp{IC} implies additional manual overhead to correct mistakes.
Nevertheless, we argue that this workload remains tolerable and represents the best reported result so far, see \Cref{mseg::table::generalization}.
Upon inspection of observed segmentation errors, we believe that this error rate could be further reduced in future work.
However, given the aforementioned uncertainties in the ground truth, which primarily affect \acp{IC} 8 to 14, a perfect segmentation appears unachievable on our dataset.

The runtime of our experiments, sometimes spanning up to five days, presented a bottleneck that prevented us from conducting more complex experiments, simply because we lacked the required resources.
We could, for example, not exhaustively test all parameters of the Betti matching loss during hyperparameter optimization.
For similar reasons, we were unable to conduct an in-depth exploration of other topology-based loss functions.

While we publish our models and scripts as open source, we are unable to do so for the datasets of 13 of the 14 different \acp{IC} for legal reasons.
Only the dataset we adopted from Rothaug \etal~\cite{DBLP:conf/ashes/RothaugKABP0P23} was published by the original authors.
In all other cases, copyright law, license agreements, and the risk of legal action by either the designers or manufacturers of the \acp{IC} result in an unfortunate legal situation that could threaten not only us as researchers but also the publishers of works like ours.

\subsection{Future Work}
\label{mseg::subsection::future_work}
While working on this publication, \ac{SAM3} has been released~\cite{carion2025sam}.
At the time of writing, the model was only available on request.
Furthermore, performance data published by Meta does not indicate a significant improvement in image segmentation; therefore, we do not expect a substantial benefit from switching to \ac{SAM3}.
Nonetheless, the applicability of \ac{SAM3} to metal line segmentation remains to be investigated.

We primarily see additional room for improvement in our generalization results.
Here, the use of additional, higher-quality datasets featuring more annotated images could help further improve generalization performance, as suggested by our findings in \Cref{mseg::subsection::generalization_splits}.
Ideally, such datasets could be made publicly available by \ac{IC} designers or manufacturers, who also possess the \ac{GDSII} ground truth corresponding to the images.

Finally, future research should investigate whether techniques to mitigate sample preparation or imaging defects~\cite{DBLP:conf/icip/HuangCYLSYGW21}, as depicted in \Cref{mseg::figure::bad_images}, can be incorporated into approaches like \AppName.
To this end, larger datasets containing more erroneous \ac{SEM} images, paired with reliable ground truth, would be required.

\subsection{Conclusion}
\label{mseg::subsection::conclude}
In this work, we introduced \AppName, a tool for \ac{IC} metal layer image segmentation based on \ac{SAM2}.
We demonstrated that our approach produces reliable segmentation masks with an \ac{ESD} error rate of \SI{0.62}{\percent}.
\AppName works almost flawlessly for \acp{IC} that it has been fine-tuned on, significantly improving upon existing work.
Furthermore, it shows promising results even for \acp{IC} that it was not previously exposed to.
Therefore, we have shown that \AppName generalizes well across \ac{IC} technology nodes and image capturing techniques.
However, we also see that our approach could be further enhanced by fine-tuning on additional image datasets from more diverse \acp{IC}.
Therefore, we strongly advocate for \ac{IC} designers and manufacturers to make annotated image datasets publicly available as benchmarks to improve not only the reliability and accuracy of \ac{IC} verification tools like ours, but also the reproducibility and comparability of research results.
By publishing our final \AppName model trained on images from all 14 \acp{IC}, along with our training, inference, and evaluation scripts, as open source, we take a first step in this direction and hope to lower the barriers for researchers to enter this domain.

\section*{Acknowledgements}
Funded by the Deutsche Forschungsgemeinschaft (DFG, German Research Foundation) under Germany's Excellence Strategy - EXC 2092 CASA - 390781972.


\bibliographystyle{alpha}
\bibliography{main.bib}

\end{document}